\documentclass[journal]{IEEEtran}

\usepackage{amsmath,amssymb,amsfonts}
\usepackage{graphicx}
\usepackage{booktabs} 
\usepackage[ruled,vlined,linesnumbered]{algorithm2e} 
\usepackage{cite} 
\usepackage{hyperref} 
\usepackage{subfigure}
\usepackage{latexsym}
\usepackage{xcolor}

\usepackage{multicol}
\usepackage{multirow}

\usepackage{array}

\usepackage[linewidth=1pt]{mdframed}

\begin{document}


\title{Moving or Predicting? RoleAware-MAPP: A Role-Aware Transformer Framework for \underline{M}ovable \underline{A}ntenna \underline{P}osition \underline{P}rediction to Secure Wireless Communications}

\author{Wenxu Wang, Xiaowu Liu, Wei Gong, Yujia Zhao, Kaixuan Li, Qixun Zhang,~\IEEEmembership{Member,~IEEE}, Zhiyong Feng,~\IEEEmembership{Senior Member,~IEEE}, and Kan Yu,~\IEEEmembership{Member,~IEEE}
\thanks{This work is supported by the National Natural Science Foundation of China with Grant 62301076, Fundamental Research Funds for the Central Universities with Grant  24820232023YQTD01, National Natural Science Foundation of China with Grants 62341101 and 62321001, Beijing Municipal Natural Science Foundation with Grant L232003, and National Key Research and Development Program of China with Grant 2022YFB4300403. (Corresponding author: Kan Yu)}
\thanks{W. Wang, X. Liu and K. Li are with the School of Computer Science, Qufu Normal University, Rizhao, P.R. China. E-mail: \{wangwx@qfnu.edu.cn, liuxw@qfnu.edu.cn, lkx0311@126.com\}.}
\thanks{W. Gong is with Inspur Computing Technology Pty Ltd. and Shandong Key Laboratory of Advanced Computing, Jinan, P. R. China. E-mail: gongwei@inspur.com.}
\thanks{K. Yu, Y. Zhao, Q. Zhang and Z. Feng are with the Key Laboratory of Universal Wireless Communications, Ministry of Education, Beijing University of Posts and Telecommunications, Beijing, 100876, P.R. China. E-mail: \{kanyu1108@126.com, yjzhao0318@126.com, zhangqixun@bupt.edu.cn, fengzy@bupt.edu.cn\}.}
}

\markboth{IEEE Transactions on Communications,~Vol.~, No.~, 2025}%
{Shell \Wenxu Wang{\textit{et al.}}: Shortest Link Scheduling Under SINR}
\maketitle
\begin{abstract}
   Movable antenna (MA) technology provides a promising avenue for actively shaping wireless channels through dynamic antenna positioning, thereby enabling electromagnetic radiation reconstruction to enhance physical layer security (PLS). However, its practical deployment is hindered by two major challenges: the high computational complexity of real-time optimization and \emph{a critical temporal mismatch between slow mechanical movement and rapid channel variations}. Although data-driven methods have been introduced to alleviate online optimization burdens, they are still constrained by suboptimal training labels derived from conventional solvers or high sample complexity in reinforcement learning. More importantly, existing learning-based approaches often overlook communication-specific domain knowledge—particularly the asymmetric roles and adversarial interactions between legitimate users and eavesdroppers, which are fundamental to PLS. To address these issues, this paper reformulates the MA positioning problem as a predictive task and introduces RoleAware-MAPP, a novel Transformer-based framework that incorporates domain knowledge through three key components: role-aware embeddings that model user-specific intentions, physics-informed semantic features that encapsulate channel propagation characteristics, and a composite loss function that strategically prioritizes secrecy performance over mere geometric accuracy. Extensive simulations under 3GPP-compliant scenarios show that RoleAware-MAPP achieves an average secrecy rate of 0.3569 bps/Hz and a strictly positive secrecy capacity of 81.52\%, \emph{outperforming the strongest baseline by 48.4\% and 5.39 percentage points}, respectively, while maintaining robust performance across diverse user velocities and noise conditions.
   
\end{abstract}

\begin{IEEEkeywords}
	Movable antenna, position prediction, physical layer security, Transformer framework
\end{IEEEkeywords}


\section{Introduction}
With the exponential growth of network access points and data throughput in wireless environments, multi-input multi-output (MIMO) technology has emerged as a fundamental solution to meet the increasing demands for high-capacity and low-latency communications. By deploying multiple antennas and establishing multiple transceiver links, MIMO systems have successfully met the growing communication requirements~\cite{Wang2024}. However, when confronting complex interference scenarios and stringent quality-of-service demands, particularly in urban environments, traditional fixed-position antenna (FPA) systems increasingly expose their inherent limitations in terms of flexibility and performance-cost trade-offs.

Movable antenna (MA) technology has been proposed as a transformative approach that exploits spatial degrees of freedom (DoF) without increasing the number of physical antenna elements~\cite{29Zhu2024oppo}. By dynamically adjusting the antenna positions and orientations across one to six dimensions, MA systems can actively shape wireless channels rather than passively adapting to it, achieving electromagnetic radiation reconstruction (ERR)—the ability to actively reconstruct and optimize electromagnetic wave propagation patterns in real-time. This ERR capability enables significant performance improvements in channel capacity, interference suppression, and physical layer security~\cite{Zhu2024Modeling,Ma2024Multi, Feng2025Movable}. Despite these advantages, the practical implementation of MA systems faces two major challenges. First, the incorporation of antenna mobility transforms conventional optimization problems into high-dimensional dynamic non-convex formulations. The resulting computational complexity, often addressed via iterative algorithms, becomes prohibitive for real-time operation \cite{Liu2025Discrete}. Second, and more critically, a fundamental temporal mismatch exists between the slow mechanical movement of antennas (on the order of milliseconds to seconds) and the rapid dynamics of user mobility and channel variations. This mismatch leads to outdated channel state information (CSI) by the time antennas are repositioned, severely degrading system performance \cite{Wang2025Throughput}.

In response, recent studies have explored data-driven methods to circumvent online optimization. These intelligent decision-making approaches can be broadly categorized into supervised learning methods and reinforcement learning frameworks. Several machine learning (ML) models have been proposed to learn direct mappings from system states to optimal MA positions, shifting computational load to an ``offline training-online inference'' paradigm. Although such models can achieve millisecond-level inference latency, their performance is inherently bounded by the quality of training labels—typically generated via suboptimal numerical solvers—which limits achievable gains. Reinforcement learning (RL) offers an alternative by framing MA positioning as a Markov decision process \cite{34Weng2024Learning,Hung2025Reinforcement}. While RL can adaptively learn control policies through environmental interaction, it suffers from high sample complexity, training instability, and difficulties in scaling to high-dimensional state spaces. These issues hinder its applicability in real-world dynamic settings \cite{Luo2024}.

The growing integration of artificial intelligence (AI) with wireless communications opens new pathways for intelligent MA control. Recent works have begun to leverage AI for communication-specific tasks, yet a deeper incorporation of domain knowledge—particularly in physical layer security (PLS)—remains underexplored \cite{portLLM,LLM4CP}. In PLS, the distinct roles of legitimate users (Bob) and eavesdroppers (Eve) introduce asymmetric objectives rooted in game-theoretic interactions. Effectively embedding such role-aware semantics into learning frameworks is crucial for optimizing secrecy performance \cite{Huang2018Game,Wu2022Game}. These considerations motivate the central research questions addressed in this work: \emph{Can we effectively extract and leverage spatio-temporal features—including spatial geometry, temporal dynamics, and user role semantics—to enhance MA position prediction?}
More specifically, \emph{how can we construct an accurate mapping from historical MA configurations, user CSI, and location data to future optimal antenna placements, thereby enabling real-time trajectory planning?}

To tackle these challenges, this paper introduces RoleAware-MAPP, a role-aware Transformer framework designed for MA position prediction. \textcolor{black}{This framework represents a novel security-oriented adaptive control strategy that integrates intelligent decision-making with ERR capabilities.} By integrating domain-specific knowledge and explicitly modeling the asymmetric characteristics of different user roles, the proposed approach achieves high predictive accuracy while maintaining real-time inference capability. The main contributions of this paper are summarized as follows:

\begin{itemize}
	\item The paper innovatively transforms the intractable, non-convex problem of real-time MA position optimization into a tractable supervised learning task. By framing it as a predictive challenge—mapping historical channel and user data to future optimal antenna positions—this approach effectively bypasses the prohibitive computational complexity and inherent latency mismatch of traditional optimization methods, paving the way for a practical solution.
	
	\item The  RoleAware-MAPP was proposed, which is a novel Transformer-based framework specifically engineered for communication security. Its core innovations lie in the deep integration of domain knowledge, featuring a role-aware embedding mechanism that asymmetrically models legitimate users and eavesdroppers, and a communication semantic extractor that provides strong, physics-informed inductive bias. The entire model is guided by a composite loss function that prioritizes security performance over mere geometric accuracy.
	
	\item Through extensive simulations in realistic 3GPP vehicular scenarios, we demonstrate the superior performance and robustness of the proposed framework. The results validate our design philosophy: \textit{RoleAware-MAPP} significantly outperforms state-of-the-art baselines in critical security metrics, such as Average Secrecy Rate (ASR) and Strictly Positive Secrecy Capacity (SPSC), across a wide range of user velocities and noise conditions, confirming its effectiveness for dynamic wireless environments.
\end{itemize}
	
The remainder of this paper is organized as follows. Section \ref{sec:related_works} reviews the related work in movable antenna systems and physical layer security. In Section \ref{sec:system_model}, we introduce the system and channel model for our scenario and provide detailed problem formulation, transforming the intractable optimization problem into a predictive task. Section \ref{sec:Architecture} elaborates on the architecture of our proposed RoleAware-MAPP framework. We present and analyze the extensive simulation results in Section~\ref{sec:numerical_results} to validate the model's performance. Finally, Section~\ref{sec:conclusions} concludes the paper and discusses potential future research directions.
	
\emph{Notations:} In this paper, $\textbf{x}^T$ and $\textbf{x}^H$ represent the transpose and conjugate transpose of the matrix or vector $\textbf{x}$, respectively. $ \mathbb{C}^{a \times b}$ and $ \mathbb{R}^{a \times b}$ respectively denote \(a \times b\) dimensional complex matrices and \(a \times b\) dimensional real matrices. \(||\mathbf{x}||_{2}\) represents the 2-norm of the vector \(\mathbf{x}\). \(\text{tr}(\mathbf{x})\) and \(\rm{diag}(\mathbf{x})\) respectively represent the trace and the diagonal matrix with diagonal elements \(\mathbf{x}\). \(\left| \mathbf{x} \right|\) represents taking the modulus of vector \(\mathbf{x}\). \(\nabla_{x}\) and \(\frac\partial{\partial x}\) respectively denote the gradient operator and the partial derivative operator. \([x]^{+}\) represents \(\max\{x, 0\}\).

\section{Related works}\label{sec:related_works}
From the perspective of PLS, the application of MA to enhance the secrecy performance in wireless communications has garnered substantial research interest. In this section, we review the state-of-the-art advances across three key domains: ERR methods, intelligent decision-making frameworks, and security-oriented adaptive control strategies. A comparative summary of representative studies is provided in Table \ref{tab:related_work}.

\subsection{Communication Reliability Enhancement based on ERR}
MA  can enable a transition from ``passive channel adaptation'' to ``active channel construction'' by dynamically reconfiguring the positions and orientations of antenna elements~\cite{29Zhu2024oppo}. Recent works have demonstrated its effectiveness in exploiting DoF \cite{31Ma2024MIMO}, suppressing interference \cite{30yu2025secure,40Kang2025NMAP,41Tang2025Jamming}, enabling flexible beamforming \cite{34Weng2024Learning,35Xie2025Beamforming,36shao2025hybridnearfarfield6d}, modulating null-steering \cite{40Kang2025NMAP}, and enhancing the strength of received signal \cite{31Ma2024MIMO,41Tang2025Jamming}. 

\begin{table*}[t]

	\caption{\small Comparison of Representative Works on MA System Optimization}
	\label{tab:related_work}
	\centering
	\arrayrulewidth=0.5pt
	\renewcommand{\arraystretch}{1.4}
	\begin{tabular}{|
			>{ \centering\arraybackslash}m{1.0cm}|
			>{ \centering\arraybackslash}m{3.5cm}|
			>{ \centering\arraybackslash}m{1.4cm}|
			>{ \centering\arraybackslash}m{1.4cm}|
			>{ \centering\arraybackslash}m{1.4cm}|
			>{ \centering\arraybackslash}m{1.4cm}|
			>{ \centering\arraybackslash}m{1.4cm}|
			>{ \centering\arraybackslash}m{1.4cm}|
			>{ \centering\arraybackslash}m{1.4cm}|
		}
		\hline
		\textbf{Ref.} 
		& \textbf{Techniques} 
		& \textbf{Intelligent Method} 
		& \textbf{Composite Loss}
		& \textbf{Spatio-Temporal Features}
		& \textbf{Physics-Informed Features}
		& \textbf{Security-Constrained} 
		&\textbf{NMSE Metric} 
		&\textbf{Mobility}\\ \hline
		
		\cite{31Ma2024MIMO}
		& Joint Position and Covariance Matrix Optimization
		& $\times$
		& $\times$
		& $\times$ 
		& $\checkmark$
		& $\times$ 
		& $\times$
		& $\times$
		\\  \hline
		
		\cite{32Shao2025Distributed}
		& 6DMA Three-Stage CSI Estimation
		& $\times$
		& $\times$
		& $\times$ 
		& $\checkmark$
		& $\times$ 
		& $\checkmark$
		& $\times$
		\\  \hline
		
		\cite{33Ma2025RIS}
		& Joint MA-RIS Optimization
		& $\times$
		& $\times$
		& $\times$ 
		& $\checkmark$
		& $\checkmark$
		& $\times$
		& $\checkmark$
		\\  \hline
		
		\cite{34Weng2024Learning}
		& MADDPG-based Heterogeneous Multi-Agent Learning
		& $\checkmark$
		& $\times$
		& $\times$ 
		& $\times$ 
		& $\times$ 
		& $\times$
		& $\checkmark$
		\\  \hline
		
		\cite{35Xie2025Beamforming}
		& An ISAC System Empowered by MADDPG
		& $\checkmark$
		& $\times$
		& $\times$ 
		& $\times$ 
		& $\times$ 
		& $\times$
		& $\checkmark$
		\\  \hline
		
		\cite{36shao2025hybridnearfarfield6d}
		& Hybrid-field 6DMA-DRL Optimization
		& $\checkmark$
		& $\times$
		& $\checkmark$ 
		& $\checkmark$
		& $\times$ 
		& $\times$
		& $\checkmark$
		\\  \hline
		
		\cite{37zhao2025movableantennaenhancedfederated}
		& Optimization of Federated LLM Training
		& $\checkmark$
		& $\times$
		& $\times$ 
		& $\times$ 
		& $\times$ 
		& $\times$
		& $\checkmark$
		\\  \hline
		
		\cite{38bai2024movableantennaequippeduavdata}
		& DRL-driven Joint Optimization of UAV Trajectory and MA Orientation
		& $\checkmark$
		& $\times$
		& $\checkmark$
		& $\times$ 
		& $\times$ 
		& $\times$
		& $\checkmark$
		\\  \hline
		
		\cite{39Jang2025Estimation}
		& Neural Network-aided Channel Estimation
		& $\checkmark$
		& $\times$
		& $\times$ 
		& $\checkmark$
		& $\times$ 
		& $\checkmark$
		& $\checkmark$
		\\  \hline
		
		\cite{40Kang2025NMAP}
		& Adaptive Loss Function-driven Multi-beam Beamforming
		& $\checkmark$
		& $\checkmark$
		& $\times$ 
		& $\times$ 
		& $\checkmark$
		& $\times$
		& $\times$
		\\  \hline
		
		\cite{41Tang2025Jamming}
		& MLP-based MA Position Optimization for Anti-Jamming
		& $\checkmark$
		& $\times$
		& $\times$ 
		& $\times$ 
		& $\times$ 
		& $\times$
		& $\checkmark$
		\\  \hline
		
		\cite{43yu2025predictive}
		& Transformer-LSTM Forecasting Framework
		& $\checkmark$
		& $\times$
		& $\checkmark$
		& $\times$ 
		& $\checkmark$
		& $\checkmark$
		& $\checkmark$
		\\  \hline
		
		\textbf{this work} 
		& \textbf{RoleAware-MAPP}
		& $\checkmark$ 
		& $\checkmark$
		& $\checkmark$ 
		& $\checkmark$
		& $\checkmark$  
		& $\checkmark$  
		& $\checkmark$\\  \hline
		
	\end{tabular}
	\vspace{0.5cm}
	\raggedright
\end{table*}

In \cite{30yu2025secure}, considering the integrated sensing and communication (ISAC) systems, Yu \textit{et al.} identified MA as a key enabler for the control of the radiation pattern and the adaptation of the electromagnetic environment. In particular, they proposed an interference-coupling framework that jointly ensures communication reliability, security, and sensing accuracy. In \cite{31Ma2024MIMO}, Ma \textit{et al.} jointly optimized antenna positions and signal covariance matrices to maximize channel capacity, showing significant gains over conventional FPA-based MIMO systems. To address channel estimation challenges in MA setups, in \cite{32Shao2025Distributed}, Shao \textit{et al.} leveraged the directional sparsity of six-dimensional movable antennas (6DMA) and designed a three-stage CSI estimation protocol. In \cite{33Ma2025RIS}, Ma \textit{et al.} further introduced a joint optimization framework of MA and reconfigurable intelligent surface (RIS) parameters, achieving synergistic improvements in sensing, communication, and security.

\subsection{Intelligent-Assisted ERR Methods}

The integration of AI and ML, particularly deep reinforcement learning (DRL), has opened new avenues for MA system optimization. For example, in \cite{34Weng2024Learning}, Weng \textit{et al.} proposed a heterogeneous multi-agent deep deterministic policy gradient (MADDPG) framework, in which agents independently learn beamforming and mobility policies under imperfect CSI, effectively decoupling the two learning processes and mitigating performance degradation from outdated CSI.
Extending this line of work, in \cite{35Xie2025Beamforming}, Xie \textit{et al.} developed an enhanced heterogeneous MADDPG architecture with specialized agents for antenna configuration, improving both reliability and transmission efficiency through centralized training and decentralized execution. For 6DMA systems, in \cite{36shao2025hybridnearfarfield6d}, Shao \textit{et al.} introduced a mixed-field channel model and applied DRL to jointly optimize antenna positions, orientations, and beamforming.

Considering federated learning scenarios, in \cite{37zhao2025movableantennaenhancedfederated}, Niyato \textit{et al.} proposed a holistic optimization framework combining successive convex approximation and penalty dual decomposition to jointly optimize global rounds, antenna positions, and beamforming matrices. In \cite{38bai2024movableantennaequippeduavdata}, using DRL, Bai \textit{et al.} reduced computational complexity by optimizing UAV trajectories and antenna positions, significantly lowering the overhead of online computation. In \cite{39Jang2025Estimation}, Jang \textit{et al.} designed a deep neural network (DNN)-assisted channel estimation framework that jointly optimizes antenna placement and channel modeling.

\subsection{Security-Oriented Intelligent Adaptive Control Methods}
PLS in MA-enabled systems has drawn considerable interest due to the MA's inherent capability to actively shape propagation channels \cite{40Kang2025NMAP,41Tang2025Jamming,43yu2025predictive}. In \cite{40Kang2025NMAP}, Kang \textit{et al.} modeled eavesdropping scenarios as a joint optimization of multi-beamforming and antenna positioning. In particular, they introduced an adaptive loss function that maximizes the minimum beamforming gain while adaptively suppressing interference leakage, incorporating a dynamic tradeoff parameter to balance user rate maximization against information leakage.
For multi-jammer environments, in \cite{41Tang2025Jamming}, Tang \textit{et al.} formulated the problem of PLS as SINR maximization and designed a multilayer perceptron (MLP)-based deep learning model to optimize antenna positions, achieving near-optimal anti-jamming performance with low online complexity.

Addressing the temporal mismatch between mechanical MA movement and user mobility, in our previous work \cite{43yu2025predictive}, we reformulated continuous antenna positioning as a predictive task. In detail, we developed a hybrid Transformer-LSTM network that captures spatiotemporal dependencies in historical trajectories, reducing normalized mean squared error (NMSE), a popular evaluation metric for channel estimation accuracy, by 49\% at least compared to benchmark \cite{Liu2021Location,Hu2024LSTM} and improving practicality in dynamic environments.

\textbf{Motivation:} To sum up, significant developments have been made in AI-driven MA's system optimization in terms of PLS, but a key limitation remains. That is, most existing methods treat MA positioning as a generic sequence modeling problem, lacking deep integration of communication-specific domain knowledge, such as the asymmetric roles of legitimate users and Eves in PLS. Moreover, the fundamental mismatch between slow mechanical movement of MA systems and fast channel dynamics of real environments continues to challenge real-time control effectiveness. Motivated by these gaps, this paper introduces a novel predictive framework that explicitly embeds domain knowledge to overcome latency limitations and directly optimize security-centric performance.

\section{System and Channel Model}\label{sec:system_model}

\subsection{System Model}
\begin{figure*}[htbp]
    \centering
    \includegraphics[width=0.8\linewidth]{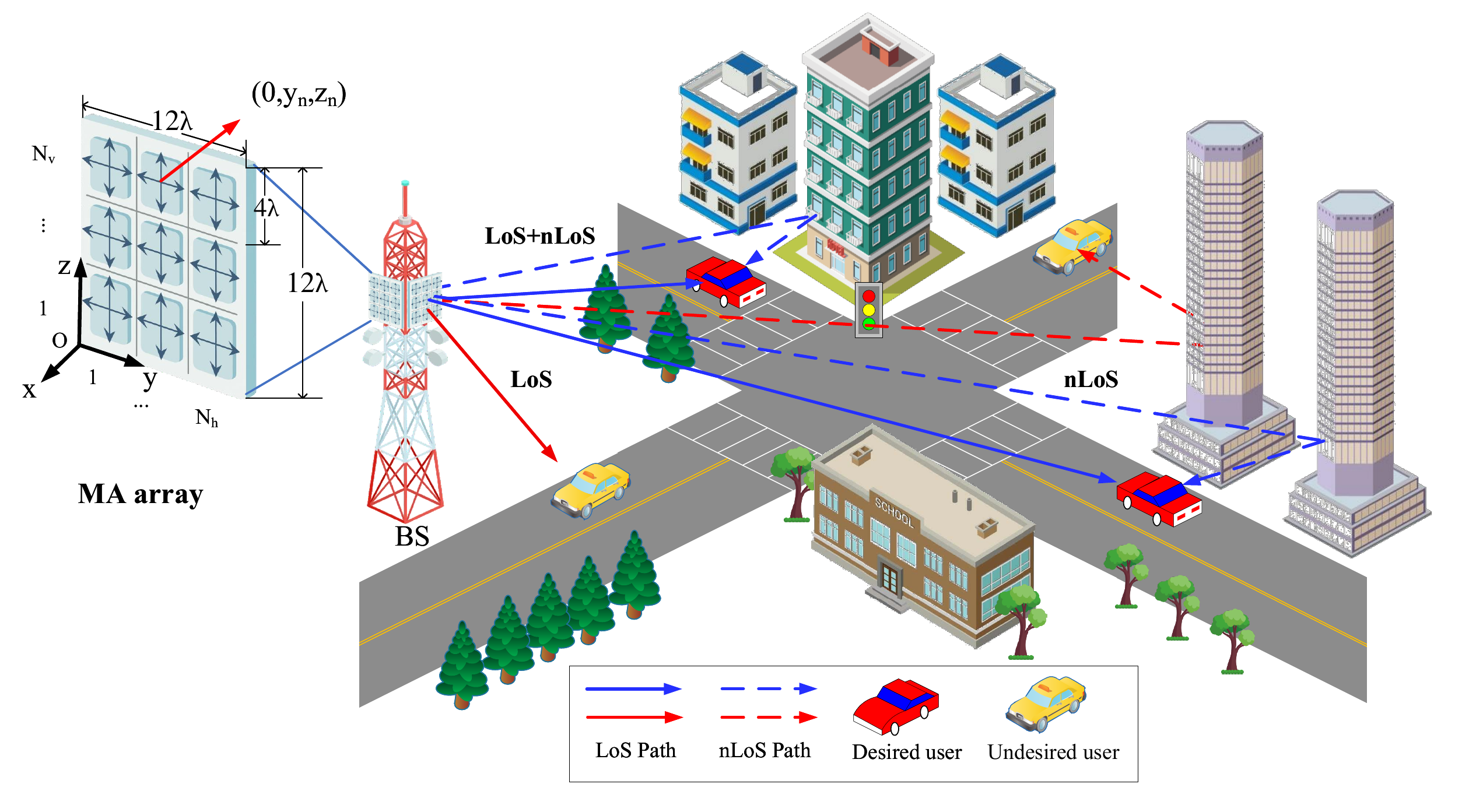} 
    \caption{System model of the MA-enabled communication scenario}
    \label{fig:system_model}
\end{figure*}

A downlink time-division duplexing (TDD) vehicular communication system is considered, as illustrated in Fig.~\ref{fig:system_model}. The base station (BS), located at a fixed position of $(0, 0, h_\text{BS})$ in a 3D Cartesian coordinate system, serves multiple users within a dynamic urban environment, where $h_\text{BS}$ represents the height of BS. The scenario is characterized by the presence of two distinct user roles: a legitimate desired user (Bob) and an illegitimate passive eavesdropper (Eve). Both users equipped with a single antenna are mobile and navigate through an environment with both line-of-sight (LoS) and non-line-of-sight (NLoS) propagation paths. Denote the 3D coordinates of the $k$-th user as $\mathbf{u}_k = [x_k, y_k, h_v]^\mathrm{T}$, where $k \in \{1, 2, \ldots, K\}$ and $h_v$ represents the vehicle height. The total number of users is $K = U + V$, where $U$ and $V$ denote the number of desired and undesired users, respectively.

The BS is equipped with a MA array, which serves as the core component of our system model. The array consists of $N_t = N_h \times N_v$ antenna elements, with $N_h$ and $N_v$ representing the number of antennas in the horizontal and vertical directions, respectively. As illustrated in Fig.~\ref{fig:system_model}, each antenna element can be dynamically repositioned within an individual square region of side length $4\lambda$, while the entire array is confined within a square aperture of dimension $D = 12\lambda \times 12\lambda$, where $\lambda$ is the carrier wavelength. The local coordinate of the $n$-th antenna element, relative to the array center at time $t$, is given by $\mathbf{p}_n(t) = [0, y_n(t), z_n(t)]$ for $n = \{1, \dots, N_t\}$. This geometric reconfigurability introduces additional degrees of freedom to actively manipulate the wireless channel.

\subsection{Channel Model}

Based on the aforementioned system model, a geometry-based multipath channel model is adopted to describe the time-varying wireless propagation in the MA-enabled vehicular environment. The channel between the $n$-th antenna of BS and the $k$-th user at time $t$ comprises one LoS path and $P$ NLoS paths. Accordingly, the channel coefficient denoted by $h_{n,k}(t)$, can be represented as the superposition of these multipath components \cite{Zhang2020Efficient}. That is,
\begin{equation}
h_{n,k}(t) = \sum_{p=1}^{P+1} \alpha_p \beta_p e^{j2\pi \frac{(\mathbf{r}_{k,p}^{\text{tx}})^T \mathbf{p}_n(t)}{\lambda}} e^{j2\pi w_p t} e^{j2\pi f \tau_p},
\end{equation}
where $\alpha_p$ and $\beta_p$ represent the complex gains of the $p$-th path. The term $w_p=(\mathbf{r}_{k,p}^{\text{tx}})^T\mathbf{v}/\lambda$ denotes the Doppler frequency shift induced by user mobility, where $\mathbf{v}$ represents the velocity vector of the user and $\lambda$ is the wavelength. The parameter $\tau_p$ denotes the path delay. The vectors $\mathbf{r}_{k,p}^{\text{tx}}$ and $\mathbf{r}_{k,p}^{\text{rx}}$ represent the spherical unit vectors at the BS and user sides, respectively, and can be given by \cite{Feng2025Movable}
\begin{equation}
	\mathbf{r}^{\text{tx}} = \begin{bmatrix}
		\sin\theta_{\text{EOD}} \cos\phi_{\text{AOD}} \\
		\sin\theta_{\text{EOD}} \sin\phi_{\text{AOD}} \\
		\cos\theta_{\text{EOD}}
	\end{bmatrix}, \quad
	\mathbf{r}^{\text{rx}} = \begin{bmatrix}
		\sin\theta_{\text{EOA}} \cos\phi_{\text{AOA}} \\
		\sin\theta_{\text{EOA}} \sin\phi_{\text{AOA}} \\
		\cos\theta_{\text{EOA}}
	\end{bmatrix},
\end{equation}
where $\theta_{\text{EOD}}$ and $\phi_{\text{AOD}}$ represent the elevation angle of departure and azimuth angle of departure, respectively, while $\theta_{\text{EOA}}$ and $\phi_{\text{AOA}}$ denote the elevation angle of arrival and azimuth angle of arrival, respectively.

The complete channel vector from the BS with $N_t$ antennas to the $k$-th user can be expressed as
\begin{equation}
\mathbf{h}_k(t) = [h_{1,k}(t), \dots, h_{N_t,k}(t)]^\mathrm{T} \in \mathbb{C}^{N_t \times 1}.
\end{equation}
The signal received at the $k$-th user, after applying a beamforming vector $\mathbf{w}(t) \in \mathbb{C}^{N_t \times 1}$ at the BS, is given by
\begin{equation}
y_k(t) = \mathbf{h}_k^\mathrm{H}(t) \mathbf{w}(t) s(t) + n_k(t),
\end{equation}
where $s(t)$ denotes the transmitted symbol with unit power, and $n_k(t) \sim \mathcal{CN}(0, \sigma^2)$ is the Additive White Gaussian Noise (AWGN). Consequently, the achievable channel capacity for user $k$ at time $t$ is represented as~\cite{LLM4CP}
\begin{equation}
C_k(t) = \log_2 \left(1 + \frac{|\mathbf{h}_k^\mathrm{H}(t) \mathbf{w}(t)|^2}{\sigma^2}\right).
\end{equation}

\subsection{Performance Metric for Evaluating Secrecy, Reliability and Accuracy}\label{sec:Metrics}

Conventional MA positioning strategies are predominantly designed for instantaneous optimization, wherein antenna configurations are computed and deployed based on the current CSI. However, such approaches suffer from a fundamental \emph{temporal mismatch} in practical deployments: iterative optimization algorithms incur substantial computational latency—ranging from hundreds of milliseconds to several seconds—while the physical movement of MA elements further introduces non-negligible actuation delay. \emph{By the time an ``optimal'' MA positioning reconfiguration is attained, the underlying CSI may have already evolved significantly due to user mobility}. This issue is particularly critical in PLS contexts, where both the Bob and the Eve experience time-varying channels. To overcome this limitation, we introduce a continuous-time MA positioning prediction framework, in which one Bob and one Eve move continuously over an extended temporal horizon. The objective shifts from optimizing performance at discrete instants to sustaining enhanced secrecy performance throughout the entire communication duration. This reformulation naturally motivates a predictive optimization paradigm: \emph{instead of reacting to outdated CSI, we seek to infer future optimal MA positioning configurations from historical channel observations, thereby inherently absorbing both computational and mechanical latencies into the predictive horizon.}

Evaluating such continuous-time MA-enabled PLS systems requires simultaneously assessing three distinct performance dimensions that cannot be captured by any single conventional metric. To this end, we adopt the following three performance indicators: (i) \textit{ASR} quantifies the time-averaged capacity advantage of Bob over Eve, measuring security throughput across the entire horizon; (ii) \textit{SPSC} measures the probability of maintaining positive secrecy rates, reflecting reliability across varying conditions; (iii) \textit{NMSE} assesses MA's geometric positioning accuracy relative to theoretical optima. Mathematically, the corresponding expressions are presented as follows.

\begin{itemize}
	\item \textbf{ASR}~\cite{Shao2023ASR}: In scenarios involving both desired and undesired users, the ASR can quantify the differential channel capacities between desired and undesired users. Over a continuous period spanning $F$ time steps, it is defined as
    \begin{equation}
        \tilde{R} = \frac{1}{F} \sum_{t=1}^{F} [C_b(t) - C_e(t)]^+,
        \label{eq:secrecy_rate}
    \end{equation}
    where $C_b(t)$ and $C_e(t)$ denote the instantaneous channel capacities of Bob and Eve at time $t$, respectively, and $[\cdot]^+$ denotes $\max\{\cdot, 0\}$. A higher ASR value indicates that the system successfully maximizes the capacity advantage of the desired user over the undesired user.
	
	\item \textbf{SPSC}~\cite{SPSC}: The SPSC measures the reliability of maintaining positive differential capacity throughout the observation period. It can be defined as
	\begin{equation}
		\text{SPSC} = \Pr\left[\tilde{R} > 0\right].
	\end{equation}
    A higher SPSC value indicates that the system can more consistently maintain favorable channel conditions for the desired user relative to the undesired user across varying channel states and mobility patterns.

    \item \textbf{NMSE}~\cite{Xia2024Transformer}: To quantify the gap between actual antenna positions and optimal antenna positions, the NMSE is usually regarded as one of standard metrics. For a given MA position configuration $\hat{\mathcal{P}}_F$ and its optimal counterpart $\mathcal{P}_F$ over a time horizon $F$, the NMSE is defined as
    \begin{equation}
	   \text{NMSE} = \mathbb{E}\left[ \frac{\|\mathcal{P}_F - \hat{\mathcal{P}}_F \|_F^2}{\|\mathcal{P}_F\|_F^2} \right],
	   \label{eq:NMSE}
    \end{equation}
    where $\|\cdot\|_F$ denotes the Frobenius norm. A lower NMSE value indicates that the antenna positioning is closer to the optimal configuration.
\end{itemize}

Together, these three complementary metrics assess the system's high secrecy rate (via ASR), its transmission reliability (via SPSC), and the geometric precision of its antenna positioning (via NMSE), thereby providing a holistic evaluation of the proposed MA-enabled PLS system.

\subsection{Problem Formulation}\label{sec:Formulation}

Based on the continuous-time MA positioning prediction framework and evaluation metrics established above, we formulate the problem of the system security in terms of PLS. The objective is to maximize the secrecy rate by jointly designing MA positions and beamforming vectors throughout the time horizon, subject to physical distance among antennas, transmit power, and mechanical latency constraints.

\begin{subequations}
\label{prob:P1}
\begin{align}
\textbf{P1:} \quad \max_{\{\mathbf{p}_n(t)\}, \mathbf{w}(t)} & \quad \tilde{R} \\ 
\textrm{s.t.} & \quad \text{C1: }|\mathbf{p}_n(t)-\mathbf{p}_{n'}(t)| \geq \lambda/2, \notag \\ 
& \quad \quad \quad \mathbf{p}_n(t) \in D, n \in [1, N_t]; \\
& \quad \text{C2: }||\mathbf{w}(t)||_2^2\leq P_\text{max}; \\
& \quad \text{C3: }\Delta t \leq \tau_\text{max},
\end{align}
\end{subequations}
where C1 enforces minimum inter-antenna spacing ($\lambda/2$) and confines antenna movement within the designated region $D$; C2 limits the transmit power; C3 constrains the antenna repositioning time to account for mechanical movement limitations.

Solving Problem \eqref{prob:P1} directly in real-time is intractable due to its non-convex nature arising from the non-linear channel model and coupled optimization variables. Standard iterative algorithms (e.g., alternating optimization, successive convex approximation) not only suffer from high computational complexity—prohibitive for meeting the stringent latency requirements—but also risk converging to poor local optima. More critically, as discussed in Section \ref{sec:Metrics}, the temporal mismatch between optimization/movement delays and rapid channel dynamics renders any solution based on current CSI obsolete upon deployment.

To circumvent these fundamental limitations, we reformulate Problem \eqref{prob:P1} as a supervised predictive learning task. Specifically, we leverage historical observations over a time window $T$, including user positions 
$$(\mathbf{U}_{\text{Bob}}[i:i+T], \mathbf{U}_{\text{Eve}}[i:i+T]),$$
their corresponding CSI 
$$(\mathbf{H}_{\text{Bob}}[i:i+T], \mathbf{H}_{\text{Eve}}[i:i+T]),$$
and previously optimized MA positions $\mathcal{P}_T$,\footnote{The historical optimal positions are pre-calculated offline for training purposes using a particle swarm optimization algorithm to solve Problem \eqref{prob:P1} without latency constraints.} to predict optimal MA positions for a future time horizon $F$. This predictive paradigm shifts the computational burden from real-time online optimization to offline model training, enabling millisecond-level inference.

To align the learning objective with communication performance rather than mere geometric accuracy, we design a composite loss function that integrates multiple considerations:
\begin{equation}
\mathcal{L}_{\text{total}} = \alpha \cdot \mathcal{L}_{\text{NMSE}} + \beta \cdot \mathcal{L}_{\tilde{R}} + \gamma \cdot \mathcal{L}_{\text{st}},
\label{eq:loss_function}
\end{equation}
where $\mathcal{L}_{\text{NMSE}}$ measures geometric positioning accuracy as defined in Eq. \eqref{eq:NMSE}; $\mathcal{L}_{\tilde{R}}$ is a differentiable surrogate loss that encourages configurations yielding high secrecy rates; $\mathcal{L}_{\text{st}}$ penalizes violations of physical constraints (e.g., minimum inter-antenna spacing in C1). The weights $\alpha$, $\beta$, and $\gamma$ are dynamically adjusted during training to emphasize different objectives at various stages—initially prioritizing geometric accuracy for rapid convergence to feasible regions, then progressively emphasizing security performance.

The optimization Problem \eqref{prob:P1} is thus reformulated as
\begin{subequations}
	\label{prob:P2}
	\begin{align}
		\textbf{P2:} \quad \min_{\Omega} & \quad \mathbb{E}[\mathcal{L}_{\text{total}}] \\
		\textrm{s.t.} & \quad \hat{\mathcal{P}}_F = f_{\Omega}(\mathbf{U}_{\text{Bob}}[1:T], \mathbf{U}_{\text{Eve}}[1:T], \notag \\
		& \quad \quad \mathbf{H}_{\text{Bob}}[1:T], \mathbf{H}_{\text{Eve}}[1:T], \mathcal{P}_T), \label{eq:mapping_revised}
	\end{align}
\end{subequations}
where $\Omega$ denotes the parameters of a neural network $f_{\Omega}$ that learns the mapping from historical observations to future optimal positions. Solving Problem \eqref{prob:P2} yields a predictive model capable of inferring near-optimal MA positions in real-time, naturally absorbing computational and mechanical delays into the prediction horizon. The architecture of $f_{\Omega}$—specifically designed to integrate communication domain knowledge—is detailed in the next section.

\section{The Proposed RoleAware-MAPP Architecture}\label{sec:Architecture}
\begin{figure*}[htbp]
	\centering

	\includegraphics[width=0.8\linewidth]{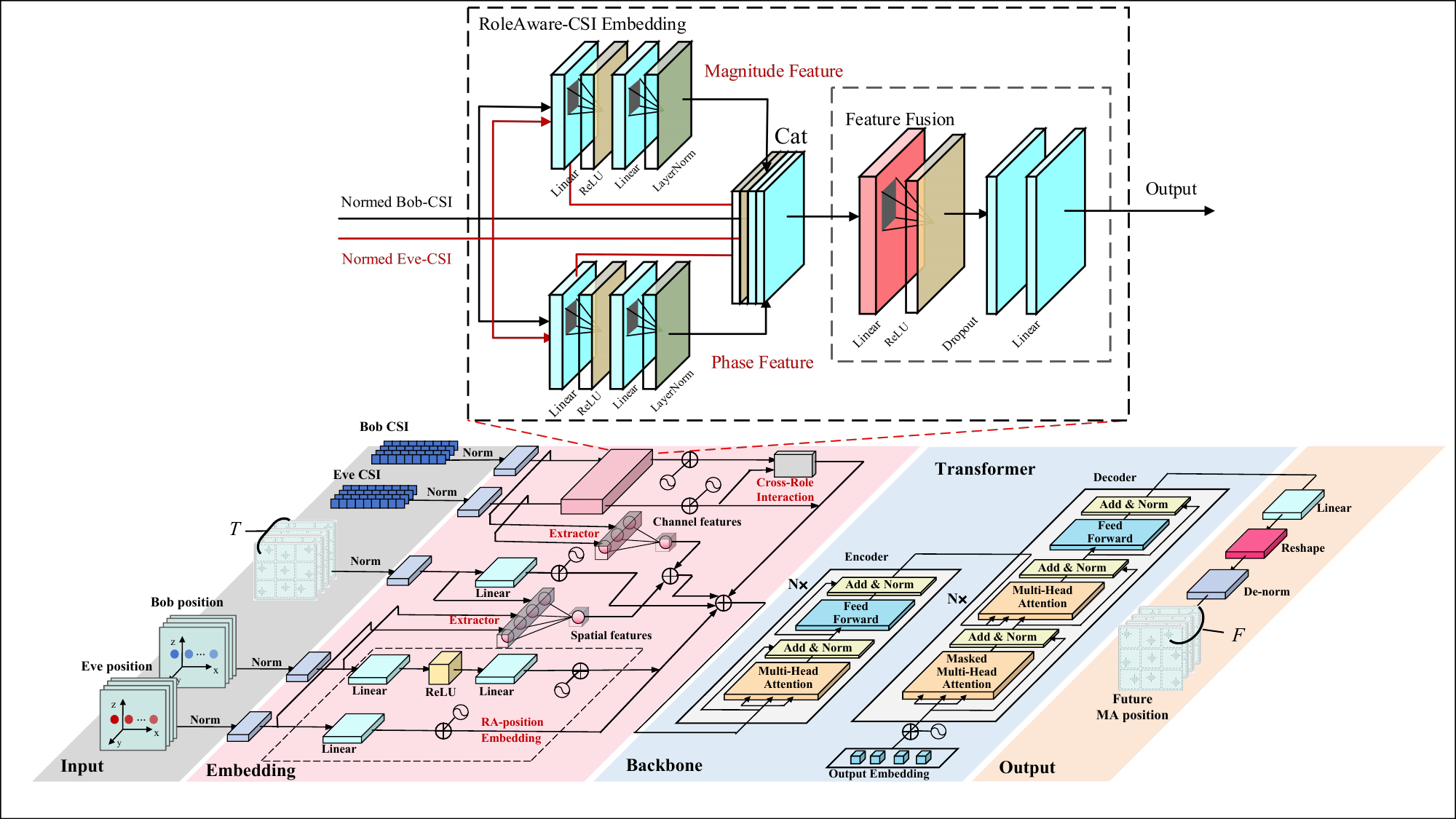} 

	\caption{The network architecture of proposed model}
	\label{fig:arc}
\end{figure*}

To address the predictive optimization problem formulated in Problem \eqref{prob:P2}, we propose RoleAware-MAPP, a novel deep learning framework built on the Transformer architecture, which is well known for its ability to capture long-range dependencies in sequential data. To tailor the model to the unique physical characteristics of wireless channels, we incorporate several domain-specific innovations. The overall framework follows a structured dataflow comprising four key components: a data preprocessor, a communication-aware embedding module, a Transformer backbone, and an output projection layer. The overall architecture is depicted in Fig.~\ref{fig:arc}. In addition, the design and rationale of each component are detailed in the following subsections.

\subsection{Data Preprocessing}
The input data first undergoes a preprocessing stage to harmonize its multi-modal nature. The model takes as input a concatenated tensor $\mathbf{X}_{\text{in}} \in \mathbb{R}^{T \times d{\text{in}}}$, where $d_{\text{in}}$ is the total input feature dimension. To address the statistical heterogeneity across the five distinct data streams, we apply a \textit{grouped normalization} strategy, where in each stream $\mathbf{X}_i$ is independently normalized using z-score standardization, namely
\begin{equation}
\bar{\mathbf{X}}_i = (\mathbf{X}_i - \mu_i) / \sigma_i,
\end{equation}
where $\mu_i$ and $\sigma_i$ denote the mean and standard deviations, respectively. The parameters $(\mu_{\mathcal{P}_T}, \sigma_{\mathcal{P}_T})$ for the MA positions are retained for the de-normalization stage.

\subsection{Communication-Aware Embedding}
This module is designed to bridge the gap between generic sequence modeling and communication-domain specificity by integrating physical-layer semantics into the representation learning process. Its architecture comprises four specialized components, each serving a distinct purpose in capturing the unique characteristics of secure MA systems: 1) role-aware feature extraction that asymmetrically models legitimate and malicious users, 2) physics-informed semantic extraction that encodes channel propagation characteristics, 3) cross-role interaction modeling that captures the adversarial relationship between communication entities, and 4) adaptive feature fusion that dynamically balances different semantic representations. This comprehensive embedding strategy enables the model to learn representations that are not only geometrically meaningful but also communication-theoretic relevant, providing a solid foundation for subsequent position prediction.

\subsubsection{Role-Aware Feature Extraction}
To reflect the asymmetric roles of Bob and Eve in physical layer security, we employ differentiated embedding pathways: Bob's data (position and CSI) are processed through deeper two-layer MLPs to capture fine-grained security-critical features, while Eve's data use simpler single-layer projections. For example, Bob's position embedding is generated as:
\begin{equation}
\mathbf{E}_{\text{Bob-pos}} = \text{LayerNorm}(\text{Linear}_2(\text{ReLU}(\text{Linear}_1(\bar{\mathbf{U}}_{\text{Bob}})))).
\end{equation}
This asymmetric design allocates more model capacity to Bob, aligning with the security objective. Similarly, we generate $\mathbf{E}_{\text{Bob-CSI}}$, $\mathbf{E}_{\text{Eve-pos}}$, $\mathbf{E}_{\text{Eve-CSI}}$, and $\mathbf{E}_{\text{MA-pos}}$ from respective input streams, all in $\mathbb{R}^{T \times d_{\text{model}}}$.

\subsubsection{Communication Semantic Extractor}
This component extracts physics-informed features directly from normalized inputs, including spatial features (user-to-BS distances, angles and velocity) and channel features (channel capacity and instantaneous secrecy rate). These features are projected into a semantic embedding $\mathbf{E}_{\text{sem}} \in \mathbb{R}^{T \times d_{\text{model}}}$ via a two-layer MLP, providing strong inductive biases grounded in communication theory.

\subsubsection{Cross-Role Interaction}
To model the adversarial relationship between Bob and Eve, we apply cross-attention where Bob's CSI queries attend to Eve's CSI keys/values:
\begin{equation}
\mathbf{E}_{\text{Bob-enh}} = \text{LayerNorm}\left(\mathbf{E}_{\text{Bob-CSI}} + \text{softmax}\left(\frac{\mathbf{Q}\mathbf{K}^\mathrm{T}}{\sqrt{d_k}}\right)\mathbf{V}\right),
\end{equation}
where $\mathbf{Q} = \mathbf{E}_{\text{Bob-CSI}}\mathbf{W}_Q$, $\mathbf{K} = \mathbf{E}_{\text{Eve-CSI}}\mathbf{W}_K$, $\mathbf{V} = \mathbf{E}_{\text{Eve-CSI}}\mathbf{W}_V$, and $\mathbf{W}_Q, \mathbf{W}_K, \mathbf{W}_V \in \mathbb{R}^{d_{\text{model}} \times d_k}$ are learned projection matrices.

\subsubsection{Feature Fusion}
The six embeddings ($\mathbf{E}_{\text{Bob-pos}}$, $\mathbf{E}_{\text{Eve-pos}}$, $\mathbf{E}_{\text{MA-pos}}$, $\mathbf{E}_{\text{Bob-enh}}$, $\mathbf{E}_{\text{Eve-CSI}}$, $\mathbf{E}_{\text{sem}}$) are aggregated via learnable weighted summation:
\begin{equation}
\mathbf{E}_{\text{fused}} = \sum_{i=1}^{6} w_i \mathbf{E}_i \in \mathbb{R}^{T \times d_{\text{model}}},
\end{equation}
where $\{w_i\}$ are trainable parameters that automatically balance each embedding's contribution.

\subsection{Transformer Backbone}
The core of our model is a standard Transformer architecture with $N_{\text{enc}}$ encoder layers and $N_{\text{dec}}$ decoder layers. The encoder receives $\mathbf{E}_{\text{fused}}$ and generates a latent contextual representation $\mathbf{Z} \in \mathbb{R}^{T \times d_{\text{model}}}$. Critically, the decoder employs a \textbf{non-autoregressive} prediction paradigm, predicting the entire future sequence of length $F$ in a single forward pass. This parallel decoding approach drastically reduces inference time, which is critical for real-time applications.
\begin{equation}
	\mathbf{O} = \text{Transformer}(\mathbf{Z}) \in \mathbb{R}^{F \times 3N_t},
\end{equation}
where Transformer(·) denotes the backbone networks.

\subsection{Output Projection}
This component translates the decoder's abstract feature representation $\mathbf{O} \in \mathbb{R}^{F \times 3N_t}$ into physical antenna positions. A fully connected layer first projects the feature dimension to the physical dimension:
\begin{equation}
\hat{\mathcal{P}}_{F, \text{norm}} = \text{Linear}(\mathbf{O}) \in \mathbb{R}^{F \times 3N_t}.
\end{equation}
Subsequently, a de-normalization step converts the predictions back to physical coordinates:
\begin{equation}
\hat{\mathcal{P}}_F = \hat{\mathcal{P}}_{F, \text{norm}} \cdot \sigma_{\mathcal{P}_T} + \mu_{\mathcal{P}_T}.
\end{equation}
This ensures that the final output is directly interpretable for controlling the MA hardware.

\section{Evaluations}\label{sec:numerical_results}

In this section, we conduct extensive simulations to evaluate the performance of our proposed RoleAware-MAPP framework. We first detail the experimental setup, and subsequently present a comprehensive analysis of the simulation results.

\subsection{Experimental Setup}

\subsubsection{Datasets}
The training, validation, and testing datasets are constructed through large-scale simulations that emulate the dynamic vehicular communication scenario outlined in Section~\ref{sec:system_model}. We utilize QuaDRiGa, a widely adopted 3GPP-compliant channel generator that implements a geometry-based stochastic channel model with high realism. A principal advantage of QuaDRiGa in our context is its antenna-agnostic design, which supports the integration of arbitrary antenna configurations. This feature is essential for simulating our MA system, as it allows us to dynamically update the position of each antenna element in the array at every time snapshot, thereby faithfully capturing channel variations resulting from reconfigurable geometry~\cite{QuaDRiGa}.

The simulation operates at a millimeter-wave carrier frequency of 28~GHz under the 3GPP Urban Macro (UMa) NLoS scenario. The BS is equipped with a $3 \times 3$ MA array, while both Bob and Eve are assumed to employ a single omnidirectional antenna. To capture a diverse range of realistic user mobility patterns, we simulate multiple motion trajectories with user velocities uniformly sampled between 10 and 100~km/h. Each data sample comprises a sequence of 20 consecutive snapshots spaced in a 0.1s interval, corresponding to an input sequence length of $T=16$ and a prediction horizon of $F=4$.

A crucial step in dataset generation is the construction of ground truth labels for supervised learning. These labels represent the theoretically optimal MA positions at each time instant. To obtain them, we treat every snapshot as an independent static optimization problem and perform an exhaustive offline search. Specifically, for each snapshot, we relax the real-world latency constraint (C3 in Problem \eqref{prob:P1}) and solve the secrecy rate maximization problem using a Particle Swarm Optimization (PSO) algorithm\cite{Zhang2025PSO}. This computationally intensive procedure yields a high-quality, albeit non-causal, suboptimal solution to the non-convex problem, serving as an upper-bound performance target for that specific instant. By repeating this process across all snapshots, we construct the ground truth trajectory sequences $\mathcal{P}_T$. The final dataset consists of 43,200 unique samples, partitioned into training (70\%), validation (15\%), and testing (15\%) subsets. Key simulation parameters are summarized in Table~\ref{tab:sim_params}.

\begin{table}[htbp]
\centering
\caption{Simulation Parameters}
\label{tab:sim_params}
\begin{tabular}{ll}
\toprule
\textbf{Parameter} & \textbf{Value} \\ \midrule
Carrier Frequency & 28 GHz\textsuperscript{\cite{Feng2025Movable}} \\
Channel Model & Urban Macro (UMa), NLoS\textsuperscript{\cite{QuaDRiGa}}\\
BS Antenna Array ($N_h \times N_v$) & $3 \times 3$\textsuperscript{\cite{Feng2025Movable}}\\
User Velocity & Uniformly in [10, 100] km/h \\
Input Sequence Length ($T$) & 16\textsuperscript{\cite{LLM4CP}}\\
Prediction Sequence Length ($F$) & 4\textsuperscript{\cite{LLM4CP}}\\
Total Samples & 43,200 \\
Dataset Split (Train/Valid/Test) & 70\% / 15\% / 15\% \\ \bottomrule
\end{tabular}
\end{table}

\subsubsection{Network and Training Parameters}
The proposed model is implemented using the PyTorch framework, with its core architecture based on a Transformer model containing 3 encoder and 3 decoder layers. The key hyperparameters for the model's structure, such as the embedding dimension, number of attention heads, and dropout rate, are detailed in Table~\ref{tab:hyper_params}. These values were selected to provide a robust balance between model capacity and computational efficiency.

For the training process, we employed the Adam optimizer, a standard choice for deep learning tasks. A warm-up and cosine annealing schedule was used to dynamically adjust the learning rate, ensuring stable convergence. The model was trained for 100 epochs with a batch size of 256, using the specific parameters listed in Table~\ref{tab:hyper_params}.

\begin{table}[htbp]
\centering
\caption{Hyperparameters for RoleAware-MAPP}
\label{tab:hyper_params}
\begin{tabular}{ll}
\toprule
\textbf{Parameter} & \textbf{Value} \\ \midrule
Embedding Dimension ($d_{\text{model}}$) & 128 \\
Encoder/Decoder Layers ($N_{\text{enc}}, N_{\text{dec}}$) & 3 \\
Attention Heads & 8 \\
Feed-Forward Dimension ($d_{\text{ff}}$) & 256 \\
Dropout Rate & 0.1 \\
Optimizer & Adam \\
Learning Rate (Max) & $1 \times 10^{-4}$ \\
Batch Size & 256 \\
Epochs & 100 \\ \bottomrule
\end{tabular}
\end{table}

\subsubsection{Baseline Schemes}
The performance of the proposed RoleAware-MAPP is benchmarked against several representative methods spanning from naive approaches to state-of-the-art deep learning architectures:

\begin{itemize}
    \item \textbf{PSO}~\cite{Zhang2025PSO}: It is a population-based metaheuristic algorithm that represents the traditional iterative optimization paradigm for solving non-convex problems.
	\item \textbf{RNN}~\cite{RNN}: A vanilla RNN processes the sequence of MA positions by maintaining and updating a hidden state, thereby providing a fundamental capability for modeling temporal channel variations.

     \item \textbf{LSTM}~\cite{Hu2024LSTM}: It utilizes gating mechanisms to mitigate the vanishing gradient problem, thereby enabling more effective modeling of long-range temporal dependencies in channel state evolution.
        
	\item \textbf{GRU}~\cite{GRU}: It simplifies LSTM architecture while maintaining comparable performance, offering a balance between computational efficiency and temporal modeling capability.
	
	\item \textbf{CNN-LSTM}~\cite{Zhang2025CNNLSTM}: It combines CNN for spatial feature extraction with LSTM for temporal sequence modeling, leveraging complementary strengths of both approaches to capture spatio-temporal channel characteristics.
	
	\item \textbf{Transformer}~\cite{Xia2024Transformer}: It leverages self-attention mechanisms to process the entire input sequence in parallel. This design circumvents the sequential processing bottleneck inherent in RNN-based models and effectively captures global dependencies across the sequence.
\end{itemize}

All deep learning baselines are trained under identical conditions with same loss function (NMSE), optimizer (Adam), and training hyperparameters to ensure a fair comparison.

\subsection{Performance Evaluation and Analysis}

\subsubsection{Training Convergence}
\begin{figure}[htbp]
	\centering
	\subfigure[$\mathcal{L}_{\text{total}} $]{
		\includegraphics[width=0.22\textwidth]{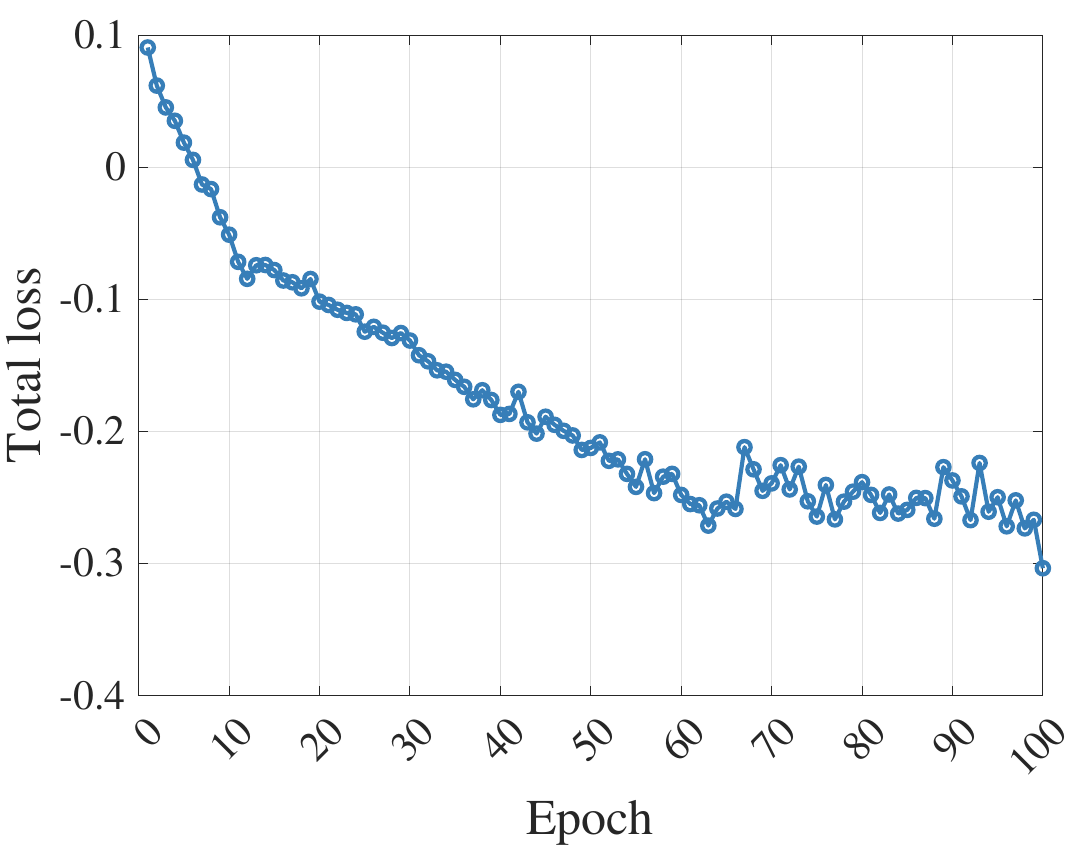}
		\label{fig:t=1}
	}
	\subfigure[$\mathcal{L}_{\text{NMSE}} $]{
		\includegraphics[width=0.22\textwidth]{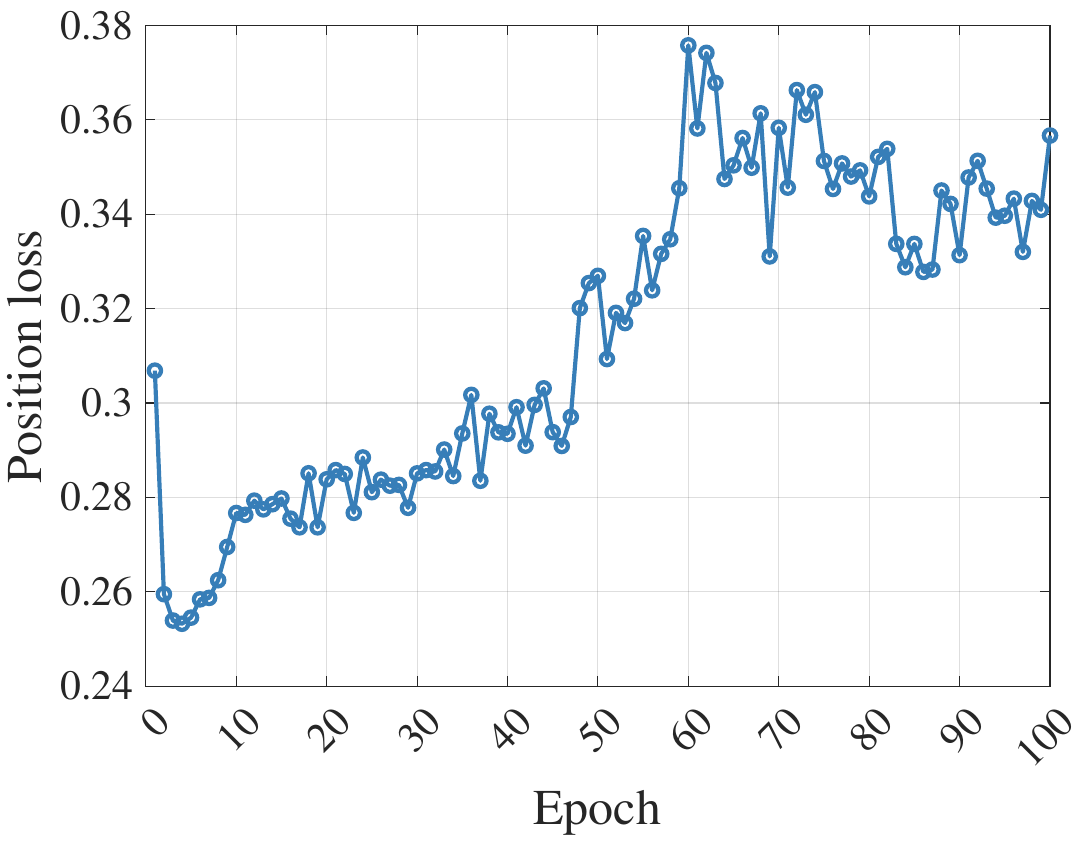}
		\label{fig:t=2}
	}
	\vspace{-0.3cm}
	\subfigure[$\mathcal{L}_{\tilde{R}}$]{
		\includegraphics[width=0.22\textwidth]{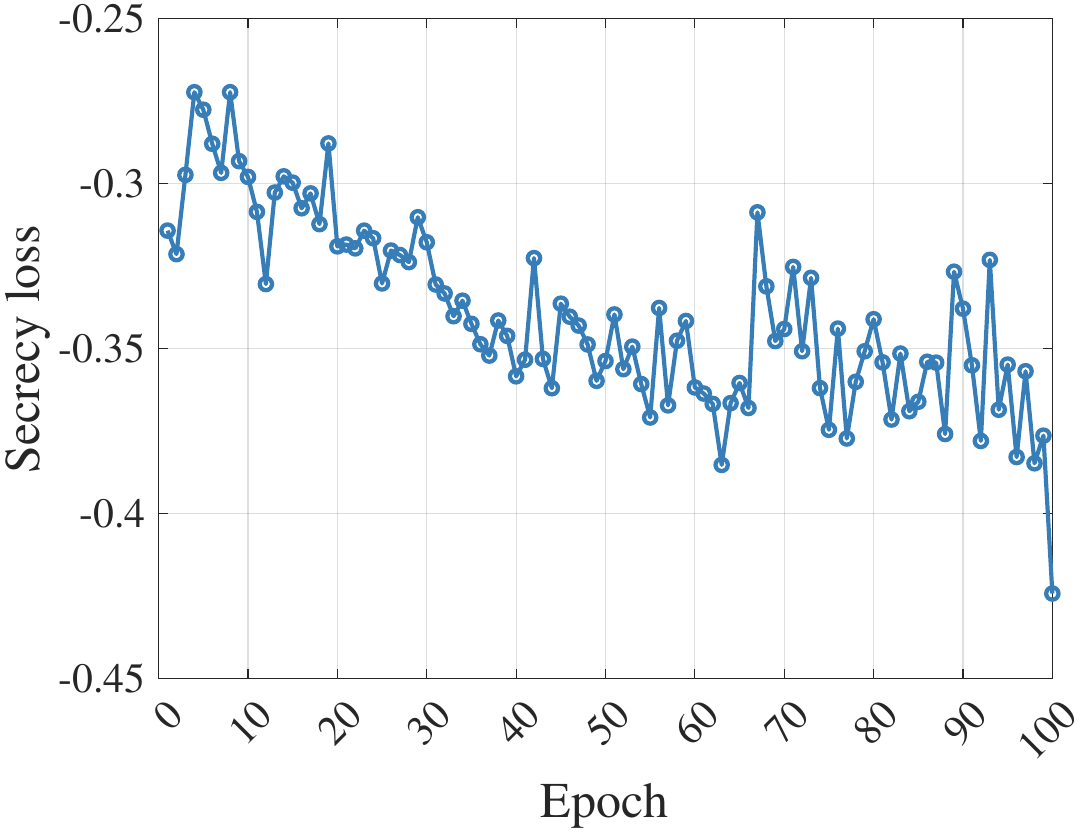}
		\label{fig:t=3}
	}
	\subfigure[$\mathcal{L}_{\text{st}} $]{
		\includegraphics[width=0.22\textwidth]{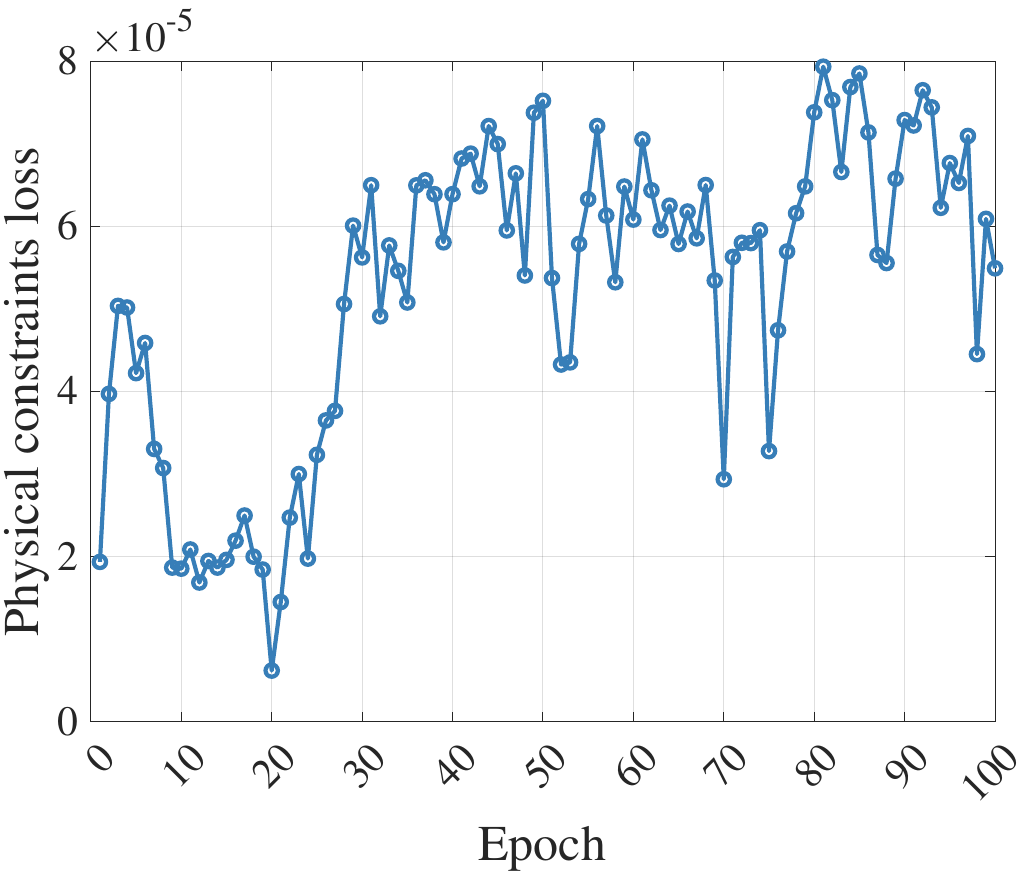}
		\label{fig:t=4}
	}

	\caption{Composite loss in training}\label{fig:CompositeLoss}
\end{figure}

Fig.~\ref{fig:CompositeLoss} depicts the progression of individual loss components throughout the 100 epochs training process of RoleAware-MAPP. As observed in Fig.~\ref{fig:t=1}, the total loss $\mathcal{L}_{\text{total}}$ \emph{follows a steady decline from an initial value of 0.1 to a final value of -0.28, indicating smooth and stable convergence. The emergence of a negative total loss is attributed to the predominance of the secrecy rate loss term $\mathcal{L}_{\tilde{R}}$}, which inherently assumes negative values when the model effectively enhances the secrecy rate—consistent with its design objective.

The training process employs a two-phase strategy with dynamic weight adjustment to guide model optimization effectively. During the initial warm-up phase (epochs 1–10), the framework prioritizes geometric accuracy by assigning a higher weight to the position prediction loss $\mathcal{L}_{\text{NMSE}}$, encouraging the model to rapidly converge toward regions near the ground truth labels. As shown in Fig.~\ref{fig:t=2}, $\mathcal{L}_{\text{NMSE}}$ decreases from 0.31 to 0.25 within the first four epochs, corresponding to a 19.4\% reduction. Subsequently, the influence of the secrecy rate loss $\mathcal{L}_{\tilde{R}}$ is progressively strengthened. From epoch 44 to 60, $\mathcal{L}_{\tilde{R}}$ rises markedly from 0.29 to 0.375—a 29.3\% increase—before stabilizing between 0.33 and 0.35, indicating the approach of a balanced trade-off with the secrecy objective.

As illustrated in Fig.~\ref{fig:t=3}, the secrecy rate loss $\mathcal{L}_{\tilde{R}}$ shows consistent improvement throughout training, declining monotonically from –0.32 to –0.43, which reflects a 34.4\% enhancement in the secrecy performance. \emph{This trend confirms the efficacy of the role-aware embedding mechanism in learning to maximize the capacity difference between the Bob and the Eve.} Meanwhile, as shown in Fig.~\ref{fig:t=4}, the physical constraint loss $\mathcal{L}_{\text{st}}$ converges rapidly to values below $2 \times 10^{-5}$ within the first 20 epochs, demonstrating the model's capability to adhere to the minimum inter-antenna spacing requirements without compromising other learning objectives.

\begin{figure}[htbp]
	\centering
	\includegraphics[width=1.0\linewidth]{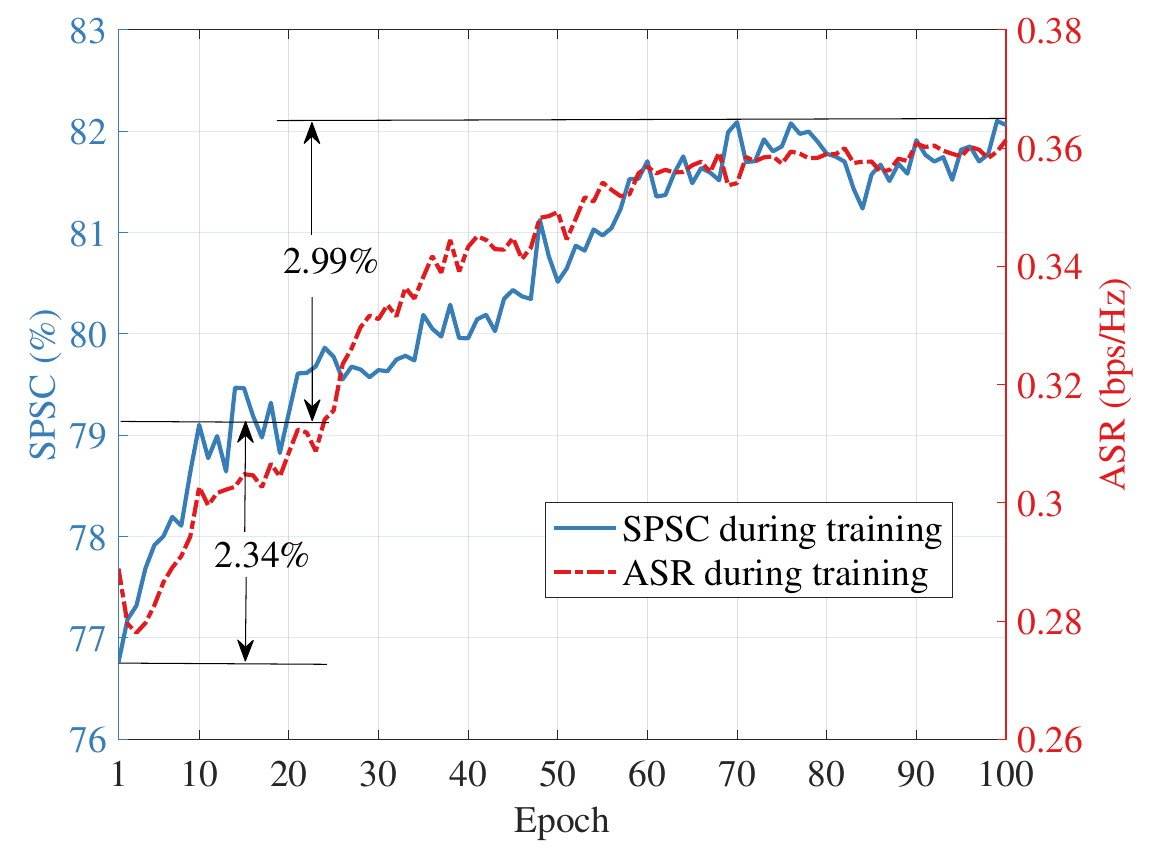} 
	\caption{SPSC and ASR of proposed model during training}\label{fig:SPSC_ASR}
\end{figure}

Next, Fig.~\ref{fig:SPSC_ASR} illustrates the influence of the two-phase training strategy on communication performance metrics. During the initial warm-up phase (epochs 1--10), the SPSC increases from 76.76\% to 79.10\%, a gain of 2.34 percentage points, as the model rapidly learns to predict antenna configurations near the ground-truth optima. In the subsequent regularization phase (epochs 11--100), where loss weights are dynamically adjusted to balance all objectives, SPSC further improves to 82.33\%, corresponding to an additional increase of 2.99 percentage points. The ASR exhibits a similar trend, rising from 0.28bps/Hz to 0.30bps/Hz during warm-up and ultimately reaching 0.36bps/Hz—a total improvement of 28.6\%.

The synchronized improvement of both metrics without significant oscillations after epoch 60 confirms the stability of the training process. The superior convergence characteristics of RoleAware-MAPP derive from three critical design elements: 1) the two-phase training strategy with warm-up ensures rapid initial convergence to feasible regions before fine-tuning for security objectives; 2) the role-aware embedding mechanism allocates differentiated model capacity between Bob and Eve, accelerating the learning of security-critical patterns; and 3) the communication semantic extractor provides physics-informed inductive bias, reducing the effective search space and preventing convergence to physically infeasible solutions.

\subsubsection{Overall Performance Comparison}
Table~\ref{tab:Overall} presents a comprehensive comparison of RoleAware-MAPP against five baseline methods across six evaluation metrics. The results demonstrate that RoleAware-MAPP achieves substantial improvements in communication-critical metrics while maintaining acceptable computational overhead.

\begin{table*}[htbp]
	\renewcommand{\arraystretch}{1.2}
	\small
	\centering
	\caption{Metrics comparison of RoleAware-MAPP and other baselines}
	\label{tab:Overall}
	\begin{tabular}{lcccccccl}
		\toprule
		\textbf{Model} & \textbf{ASR (bps/Hz)} & \textbf{SPSC} & \textbf{NMSE} & \textbf{Parameters} & \textbf{Size} & \textbf{FLOPs} & \textbf{Inference time (ms)} & \textbf{Deployment}\\ 
		\midrule
        PSO~\cite{Zhang2025PSO} & 0.2755 & 77.89\% & / & / & / & 1.73G & 3649.29 & Easy\\
		RNN~\cite{RNN} & 0.2351 & 75.39\% & 0.1491 & 79.6K & tiny & 727.58M & 0.59 & Easy\\
        GRU~\cite{GRU} & 0.2332 & 74.60\% & 0.1505 & 310.8K & small & 3.12G & 1.51  & Moderate\\
		LSTM~\cite{Hu2024LSTM} & 0.2318 & 74.25\% & 0.1568 & 277.8K & small & 2.77G & 1.16 & Moderate\\
		{CNN-LSTM}~\cite{Zhang2025CNNLSTM} & 0.2318 & 74.25\% & 0.1568 & 308.2K & small & 3.03G & 2.13  & Moderate\\
		Transformer~\cite{Xia2024Transformer} & 0.2405 & 76.13\% & \textbf{0.1475} & 673.7K & medium & 1.41G & 2.57  & Moderate\\
		RoleAware-MAPP & \textbf{0.3569} & \textbf{81.52\%} & 0.2614 & 1.32M & medium & 4.10G & 6.94 & Complex\\
		\bottomrule
	\end{tabular}
\end{table*}

In terms of ASR, the proposed RoleAware-MAPP achieves 0.3569bps/Hz, representing a 48.4\% improvement over the best baseline (Transformer model at 0.2405bps/Hz) and a 54.0\% improvement over the average baseline performance (0.2318bps/Hz). Similarly, for SPSC, RoleAware-MAPP attains 81.52\%, surpassing the Transformer baseline by 5.39 percentage points and outperforming RNN-based methods by approximately 7.27 percentage points. These significant gains in security-oriented metrics validate the effectiveness of our role-aware design philosophy.

From the perspective of NMSE, the Transformer baseline achieves 0.1475 while RoleAware-MAPP records 0.2614. This apparent disadvantage is a deliberate design choice reflecting our model's specialization. While conventional baselines optimize for geometric prediction accuracy as general-purpose sequence predictors, RoleAware-MAPP prioritizes understanding the complex relationship between antenna positions, channel states, and communication performance. The model trades marginal geometric precision for profound insights into security-critical metrics, functioning as a domain-expert system rather than a generic predictor. This specialization enables the 48.4\% ASR improvement, justifying the geometric accuracy trade-off.

Computational complexity analysis indicates that the proposed RoleAware-MAPP requires 1.32M parameters and incurs 4.10G FLOPs~\cite{flops} (floating point operations), with an average inference time of 6.94 ms per sample. Although these computational demands exceed those of baseline models—the parameter count is 1.96× that of the standard Transformer and 4.75× the average of other baselines—this investment is justified by nearly 50\% performance gains in critical security metrics. Notably, the 6.94ms inference latency remains practically feasible for MA positioning applications, where mechanical movement operates on timescales orders of magnitude larger. The substantial performance improvements achieved through this computational investment demonstrate a favorable trade-off, as the additional resources directly translate to enhanced physical layer security capabilities.

\begin{figure}[htbp]
	\centering
	\includegraphics[width=0.9\linewidth]{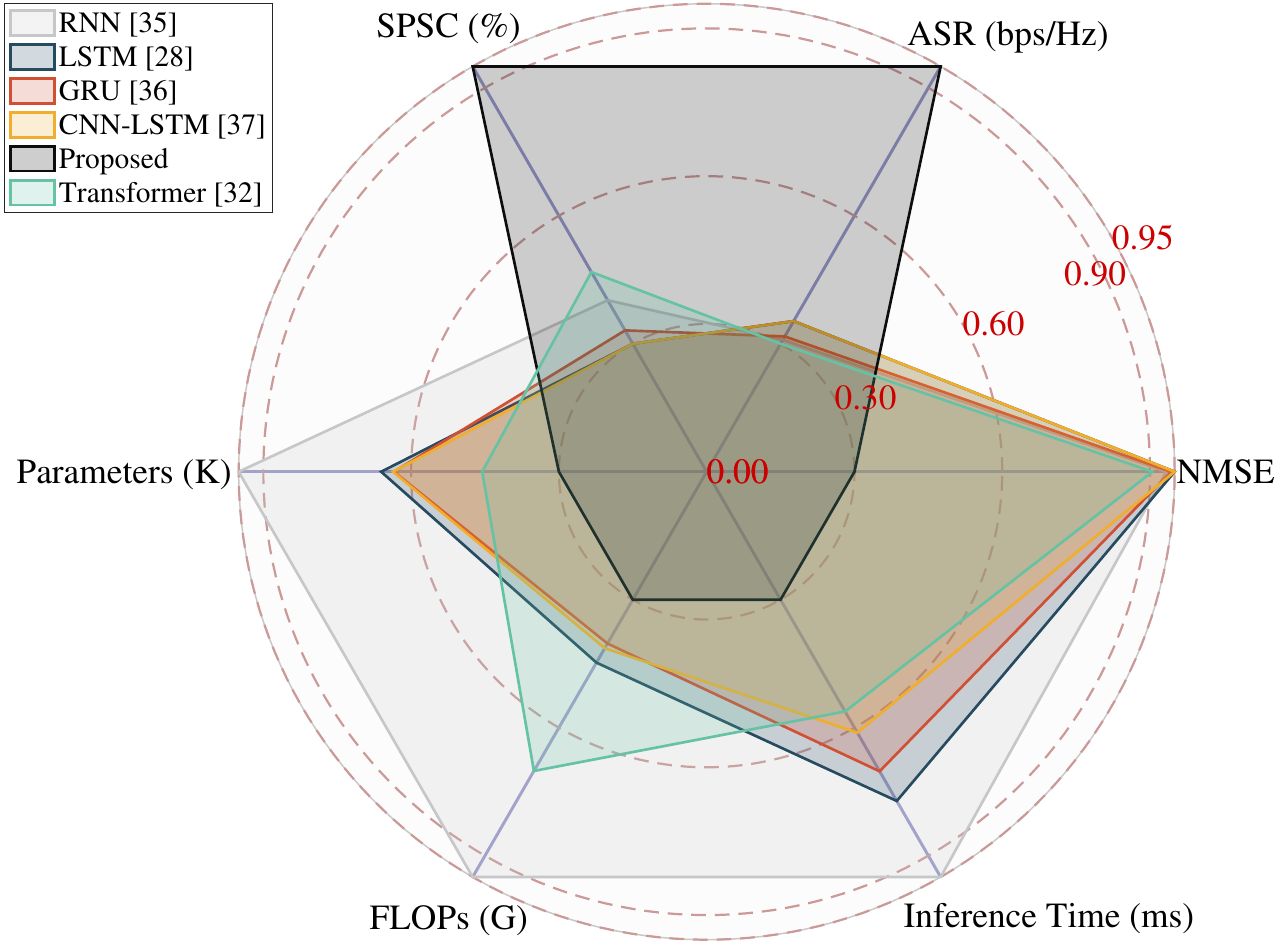} 
	\caption{Comparison of performance metrics between different models}
	\label{fig:model_compare_radar}
\end{figure}

Fig.~\ref{fig:model_compare_radar} further visualizes this performance trade-off through a radar chart, clearly illustrating RoleAware-MAPP's dominance in ASR and SPSC dimensions despite higher NMSE. The chart confirms that the proposed model successfully reallocates optimization focus from pure geometric accuracy to communication security objectives.

\begin{figure}[htbp]
	\centering
	\includegraphics[width=0.9\linewidth]{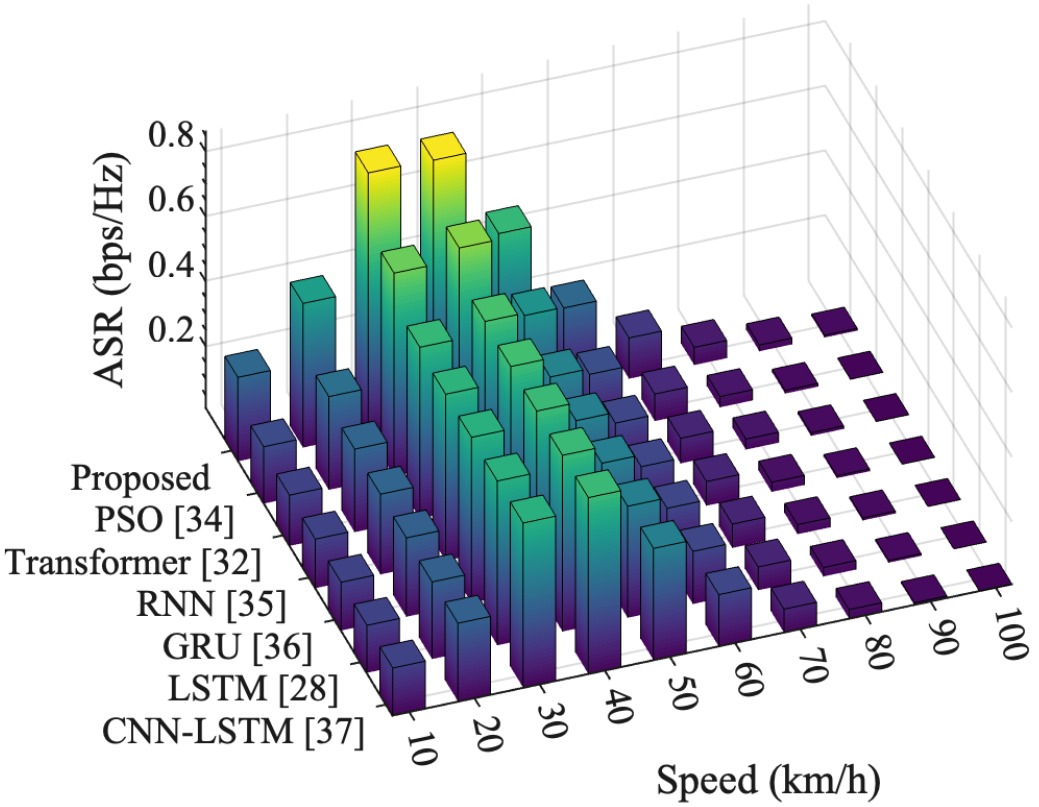} 
	\caption{ASR \textit{vs.} speed varying from 10 to 100 km/h}
	\label{fig:Speed_ASR}
\end{figure}

\begin{figure}[htbp]
	\centering
	\includegraphics[width=0.9\linewidth]{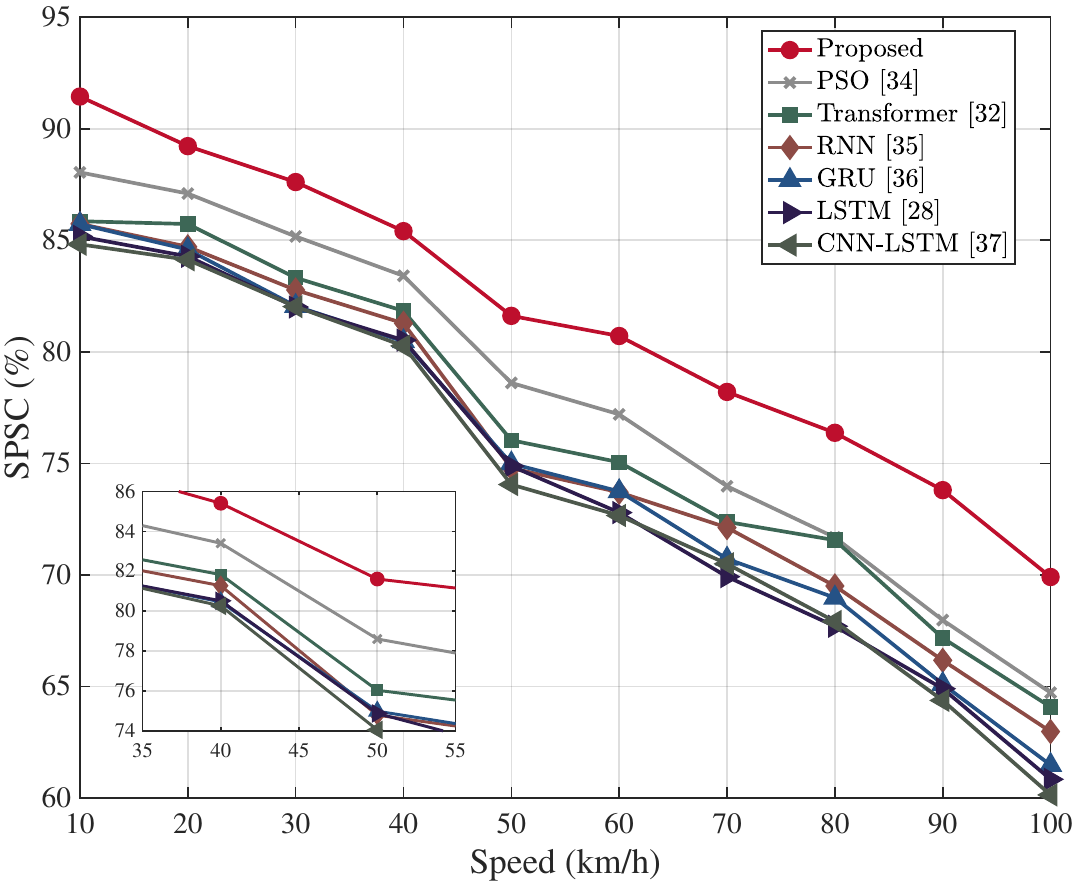} 
	\caption{ SPSC \textit{vs.} speed varying from 10 to 100 km/h}
	\label{fig:Speed_SPSC}
\end{figure}

Fig.~\ref{fig:Speed_ASR} and Fig.~\ref{fig:Speed_SPSC} evaluate model robustness across different user velocities ranging from 10 to 100km/h. For ASR performance, as given in Fig.~\ref{fig:Speed_ASR}, RoleAware-MAPP maintains superiority across all velocity ranges, achieving peak performance of 0.82bps/Hz at 30km/h while sustaining above 0.08bps/Hz even at 80km/h. In contrast, baseline methods exhibit more severe degradation, with Transformer dropping from 0.55bps/Hz to below 0.1bps/Hz. The performance gap between RoleAware-MAPP and baselines widens at higher velocities, demonstrating enhanced robustness to channel dynamics.

As shown in Fig.~\ref{fig:Speed_SPSC}, the SPSC performance demonstrates that the proposed RoleAware-MAPP maintains a success rate exceeding 70\% even at 100km/h, whereas all baseline methods drop below 65\%. Across the evaluated speed range of [10, 100]km/h, the proposed framework sustains a performance advantage of over 3\%, peaking at a margin of 7.40\% at 60km/h. This robustness is attributed to two key factors: the communication-aware embedding module effectively encodes velocity-sensitive channel features, while the cross-role interaction mechanism dynamically adapts to the evolving relationship between Bob and Eve. The consistent superiority across mobility scenarios confirms that RoleAware-MAPP is not only optimized for static configurations but also generalizes robustly to highly dynamic settings, underscoring its practical viability in real-world vehicular communication systems.

\subsubsection{Robustness Analysis}
To evaluate the robustness of RoleAware-MAPP under varying noise conditions, we examine the model performance across different background noise power levels ranging from 0dB (ideal condition) to 25dB (severe noise interference). This analysis simulates realistic scenarios where environmental noise, interference, and channel impairments degrade communication quality.

\begin{figure}[htbp]
	\centering
	\includegraphics[width=0.9\linewidth]{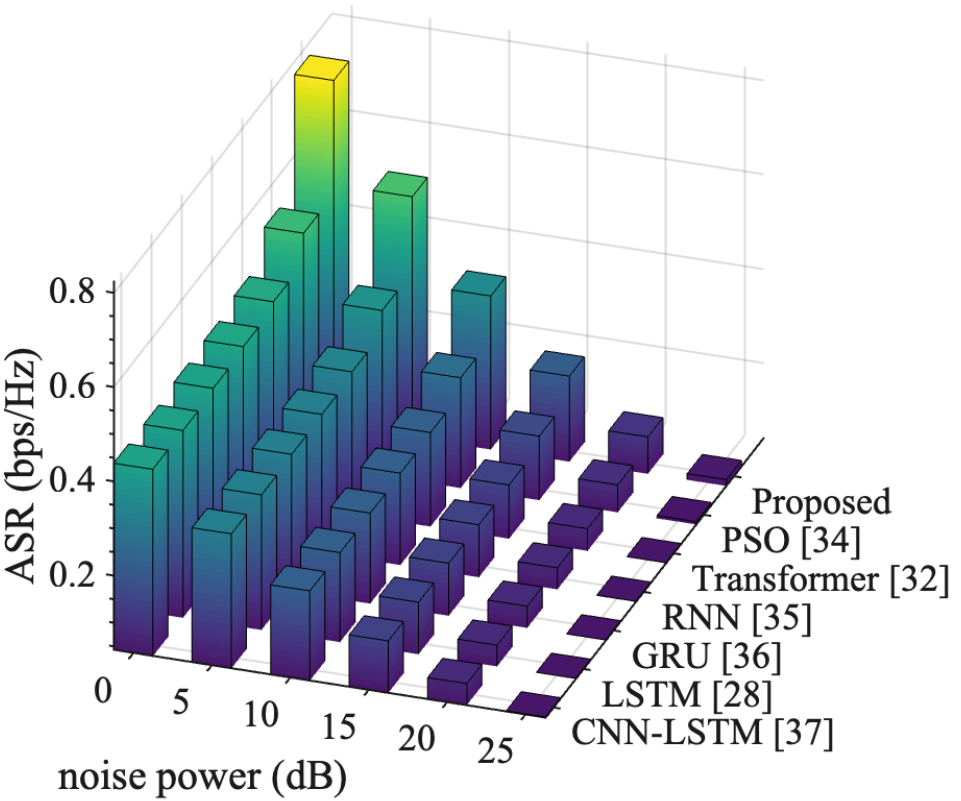} 
	\caption{ASR \textit{vs.} the level of noise power (dB) }
	\label{fig:SNR_ASR}
\end{figure}

Fig.~\ref{fig:SNR_ASR} illustrates the ASR performance degradation as noise power increases. Under ideal conditions (0dB), RoleAware-MAPP achieves 0.77bps/Hz, significantly outperforming the Transformer baseline at 0.47bps/Hz and RNN-based methods at approximately 0.45bps/Hz. As noise power increases to 25dB, all methods experience performance degradation, but RoleAware-MAPP demonstrates superior noise resilience. At 25dB noise power, RoleAware-MAPP maintains 0.055bps/Hz while baseline methods approach near-zero rates, with Transformer at 0.042bps/Hz.

\begin{figure}[htbp]
	\centering
	\includegraphics[width=0.9\linewidth]{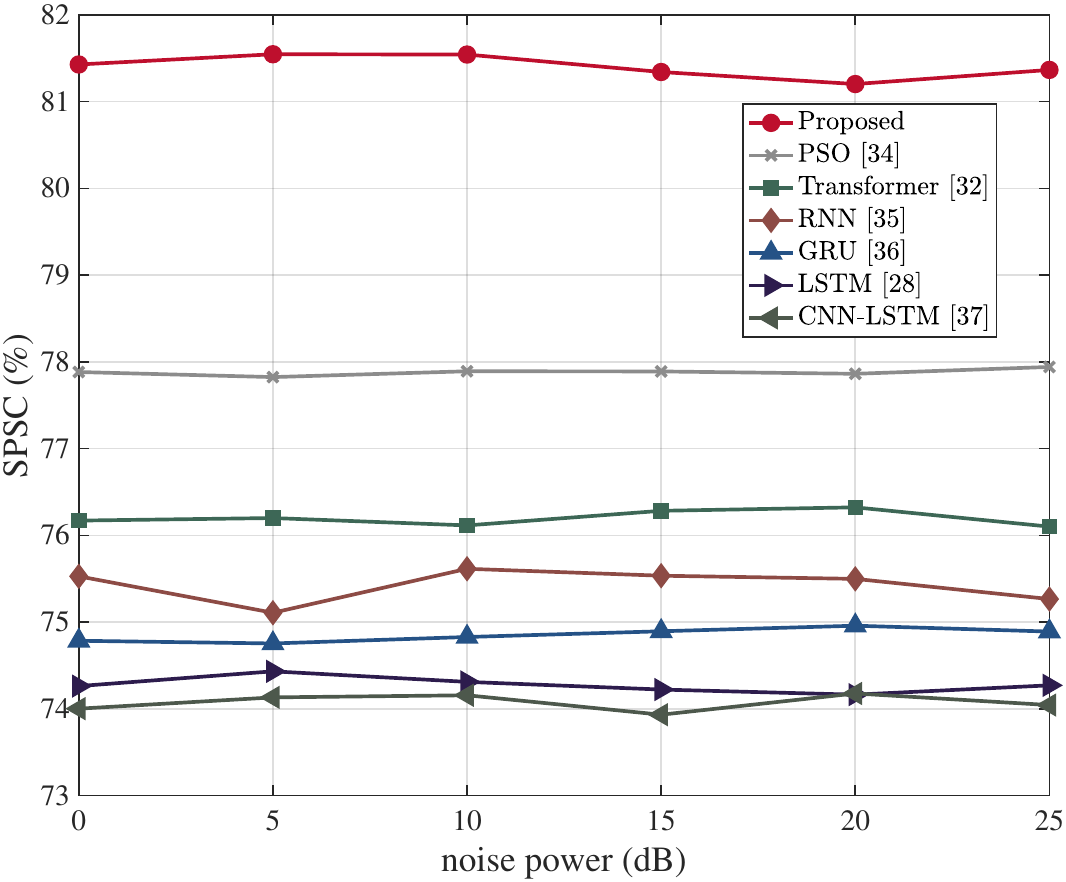} 
	\caption{SPSC \textit{vs.} the level of noise power (dB)}
	\label{fig:SNR_SPSC}
\end{figure}

Fig.~\ref{fig:SNR_SPSC} demonstrates notable consistency in SPSC performance across varying noise levels. RoleAware-MAPP sustains an SPSC of approximately 81.5\% throughout the evaluated noise power range, with fluctuations constrained within 0.3 percentage points. Similarly, baseline methods exhibit minimal performance degradation with increasing noise power: Transformer stabilizes around 76.5\%, followed by RNN (75.4\%), GRU (74.6\%), and CNN-LSTM (74.3\%). The most pronounced variation is observed in LSTM between 20dB and 25dB, though this deviation remains limited to 0.48 percentage points. This consistent performance across all models indicates their inherent capability to maintain reliable secure communication links under diverse noise conditions, with RoleAware-MAPP consistently operating at a 5–7 percentage point advantage over the competing approaches.

\subsubsection{Ablation Study}
To validate the contribution of each proposed component, we conduct comprehensive ablation experiments by systematically removing key modules from RoleAware-MAPP. Table~\ref{tab:ablation} presents the performance of three variants compared to the complete model.

\begin{table}[htbp]
	\renewcommand{\arraystretch}{1.2}
	\centering
	\caption{Results of ablation experiments}
	\label{tab:ablation}
	\begin{tabular}{lccc}
		\toprule
		\textbf{Configuration} & \textbf{ASR (bps/Hz)} & \textbf{SPSC} & \textbf{NMSE} \\ \midrule
		\textbf{RoleAware-MAPP*} & \textbf{0.3654} & \textbf{83.53\%} & 0.3100 \\
		w/o Role-Awareness & 0.3510 & 83.28\% & 0.3487 \\
		w/o Semantic Extractor & 0.3570 & 82.14\% & 0.2940 \\
		w/o composite loss function & 0.2460 & 77.08\% & \textbf{0.1498} \\ \bottomrule
	\end{tabular}
\end{table}

The most critical component is the composite loss function. When replaced with standard NMSE loss alone (w/o composite loss function), ASR dramatically decreases from 0.3654 to 0.2460bps/Hz, a 32.7\% reduction, while SPSC drops from 83.53\% to 77.08\%, losing 6.45 percentage points. Interestingly, this variant achieves the best NMSE of 0.1498, improving 51.6\% over the full model's 0.3100. This stark contrast validates our design philosophy: optimizing solely for geometric accuracy (i.e., NMSE) fails to capture the complex relationship between antenna positions and communication security. The composite loss function successfully redirects optimization focus toward security-critical objectives, justifying the geometric accuracy trade-off for substantial gains in ASR and SPSC.

Removing the communication semantic extractor (w/o Semantic Extractor) results in ASR declining to 0.3570 bps/Hz (2.3\% reduction) and SPSC dropping to 82.14\% (1.39 percentage points loss). The NMSE improves slightly to 0.2940, indicating that without physics-informed features, the model relies more heavily on geometric patterns. The semantic extractor's contribution lies in providing domain-specific inductive bias—spatial relationships, channel quality indicators, and physical constraints—that guide the model toward communication-optimal solutions rather than geometrically accurate but communication-suboptimal configurations.

The role-aware embedding mechanism (w/o Role-Awareness) contributes to performance through asymmetric feature processing. Its removal causes ASR to decrease to 0.3510bps/Hz (3.9\% reduction) and SPSC to 83.28\% (0.25 percentage points loss), while NMSE deteriorates to 0.3487. Without role differentiation, the model treats Bob and Eve symmetrically, losing the ability to explicitly maximize their capacity difference. The degraded NMSE suggests that role-aware processing not only improves security metrics, but also helps the model better understand the overall prediction task by providing clearer optimization objectives.

The ablation results reveal a clear hierarchy of component importance: the composite loss function is fundamental (32.7\% ASR impact), followed by the role-aware embedding (3.9\% impact), and the semantic extractor (2.3\% impact). However, the components exhibit synergistic effects—their combined contribution exceeds the sum of individual impacts. The full model achieves 48.7\% higher ASR than the variant without composite loss (0.3654 vs. 0.2460), demonstrating that the role-aware embedding and semantic extractor amplify the benefits of the composite loss function.

\subsubsection{Generalization Capability}
To evaluate the generalization ability of RoleAware-MAPP beyond the training configuration, we test all models with different temporal horizons: reducing the input sequence length to $T=10$ while extending the prediction horizon to $F=10$. This configuration presents a more challenging scenario—shorter historical context but longer future prediction—testing whether the learned representations transfer effectively to different temporal scales.

\begin{figure}[htbp]
	\centering
	\includegraphics[width=0.9\linewidth]{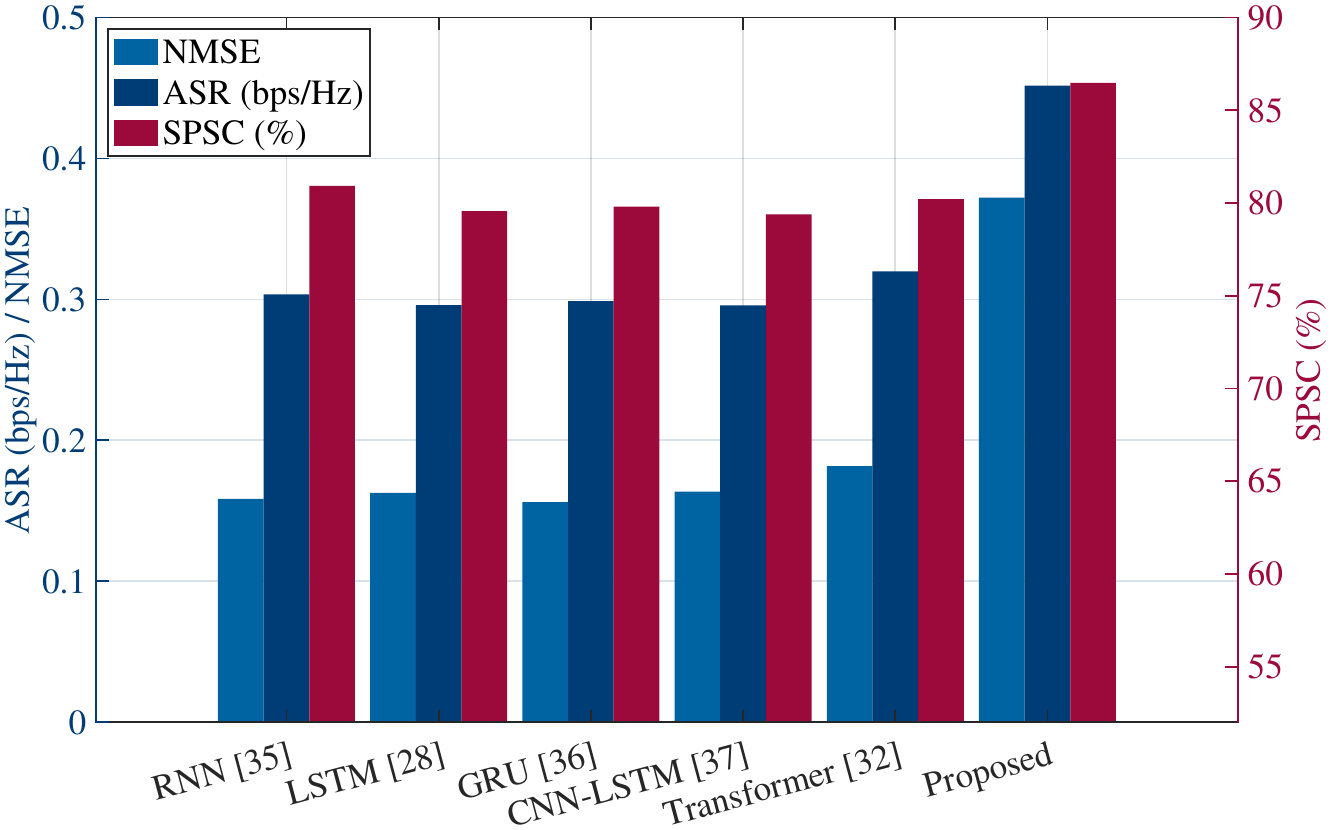} 
	\caption{Metrics comparison of models when $T=10$ and $F=10$}
	\label{fig:Generalization}
\end{figure}

Fig.~\ref{fig:Generalization} visualizes the comprehensive performance comparison, revealing that RoleAware-MAPP maintains its dominant position in ASR and SPSC metrics despite the configuration change. The performance hierarchy among methods remains consistent: RoleAware-MAPP leads significantly, followed by Transformer, with RNN-based methods trailing. However, the performance gaps widen under the new configuration, indicating that RoleAware-MAPP's specialized architecture better handles temporal scale variations.
The robust generalization validates that RoleAware-MAPP has learned transferable representations rather than overfitting to training settings, demonstrating its practical applicability when prediction requirements vary dynamically based on operational constraints.

\section{Conclusions and future works} \label{sec:conclusions}
In this paper, we tackled the critical challenges of real-time optimization and temporal mismatch in movable antenna systems by reformulating antenna positioning as a predictive learning task. To this end, we introduced RoleAware-MAPP, a novel Transformer-based framework that incorporates communication-domain knowledge through role-aware embeddings and a security-driven composite loss function, effectively prioritizing secrecy performance over geometric precision. Extensive simulations under realistic 3GPP scenarios demonstrate that the proposed framework achieves an Average Secrecy Rate of 0.3569bps/Hz and a Strictly Positive Secrecy Capacity of 81.52\%, outperforming the best-performing baseline by 48.4\% and 5.39 percentage points, respectively. These results confirm the robustness and generalization capability of RoleAware-MAPP across diverse mobility and noise conditions, underscoring its practical relevance in dynamic wireless environments. Looking forward, we plan to explore hybrid learning strategies—such as reinforcement learning—to alleviate the reliance on computationally intensive offline labels. Furthermore, implementation and validation on a physical MA testbed will be essential to assess the framework's performance under real-world impairments and hardware constraints.

\bibliographystyle{IEEEtran}
\bibliography{ref}

\end{document}